\documentclass[journal]{IEEEtran}

\usepackage{amsmath}
\usepackage{amssymb}
\usepackage{setspace}
\usepackage{epsfig}
\usepackage{epstopdf}
\usepackage{rotating}
\usepackage{caption}
\usepackage{color}
\usepackage{colortbl}
\usepackage{graphicx}
\usepackage{ifthen}
\usepackage{alltt}
\usepackage{float}
\usepackage{fancyhdr}
\usepackage{verbatim}
\usepackage{moreverb}
\usepackage{cite}
\usepackage{multirow}
\usepackage{bigstrut}
\usepackage[normalem]{ulem}

\usepackage{graphicx,subfigure,epsfig}

\usepackage{amsmath}
\usepackage{color}
\usepackage{bm}

\def\e{\begin{equation}}
\def\f{\end{equation}}
\def\=#1{\overline{\overline #1}}

\def\_#1{{\bf #1}}
\def\.{\cdot}
\def\l#1{\label{eq:#1}}
\def\r#1{(\ref{eq:#1})}

\renewcommand {\Re}{\mathop\mathrm{Re}\nolimits}

\begin{document}

\title{On-Site Wireless Power Generation}

\author{Y.~Ra'di, B.~Chowkwale, C.~A.~Valagiannopoulos,~\IEEEmembership{Senior Member,~IEEE}, F.~Liu, A.~Al\`{u},~\IEEEmembership{Fellow,~IEEE}, C.~R.~Simovski, and S.~A.~Tretyakov,~\IEEEmembership{Fellow,~IEEE}
\thanks{Y.~Ra'di and A. Al\`{u} are with the Department of Electrical and Computer Engineering,
University of Texas at Austin, Austin, TX 78712 USA (e-mail: younes.radi@utexas.edu).  A. Al\`{u} is also with the Advanced Science Research Center, City University of New York, New York, NY 10031, USA.

C.A. Valagiannopoulos is with the Department of Physics, School of Science and
Technology, Nazarbayev University, Astana, KZ-010000, Kazakhstan.

B.~Chowkwale, F.~Liu, C.~R.~Simovski, and S.~A.~Tretyakov are with the Department of Electronics and Nanoengineering, School of Electrical Engineering, Aalto University, P.~O.~Box~15500, FI-00076 Aalto, Finland.}
}


\markboth{Ra'di \MakeLowercase{\textit{et al.}}: On-Site Wireless Power Generation}%
{Ra'di \MakeLowercase{\textit{et al.}}: On-Site Wireless Power Generation}
\maketitle


\begin{abstract}
Conventional wireless power transfer systems consist of a microwave power generator and a microwave power receiver separated by some distance. To realize efficient power transfer, the system is typically brought to resonance, and the coupled-antenna mode is optimized to reduce radiation into the surrounding space. In this scheme, any modification of the receiver position or of its electromagnetic properties results in the necessity of dynamically tuning the whole system to restore the resonant matching condition. It implies poor robustness to
the receiver location and load impedance, as well as additional energy consumption in the control 
network. In this study, we introduce a new paradigm for wireless power delivery based on which the whole system, including transmitter and receiver and the space in between, forms a unified microwave power generator. In our proposed scenario the load itself becomes part of the generator. Microwave oscillations
are created directly at the receiver location, eliminating the
need for dynamical tuning of the system within the range of the self-oscillation regime. The proposed concept has relevant connections with the recent interest in parity-time symmetric systems, in which balanced loss and gain distributions enable unusual electromagnetic responses.
\end{abstract}



\section{Introduction}

The practical importance of wireless power transfer (WPT) systems has been obvious since the time of N.~Tesla, and recent advances in this field (see e.g. \cite{Science,Bred1,Bred,rev,book}), accompanied by the exponential growth in the request for fast and efficient wireless charging of battery-powered devices, has made these systems especially topical. In  conventional WPT systems, the first stage of wireless power transport from the source to the load  is a microwave generator, which transforms DC or 50/60 Hz power into microwave oscillations. This power transformation is necessary because wireless power links can operate efficiently only at reasonably high frequencies \cite{Science,Bred1,Bred,book}. Next, the microwave power available from this generator is sent  into space using some kind of antenna. This energy is received by another antenna at the receiving end, and finally converted back to the DC or 50/60 Hz form and used there. 


\begin{figure}[h!]
\centering
\subfigure[]
{\includegraphics[width=0.24\textwidth]{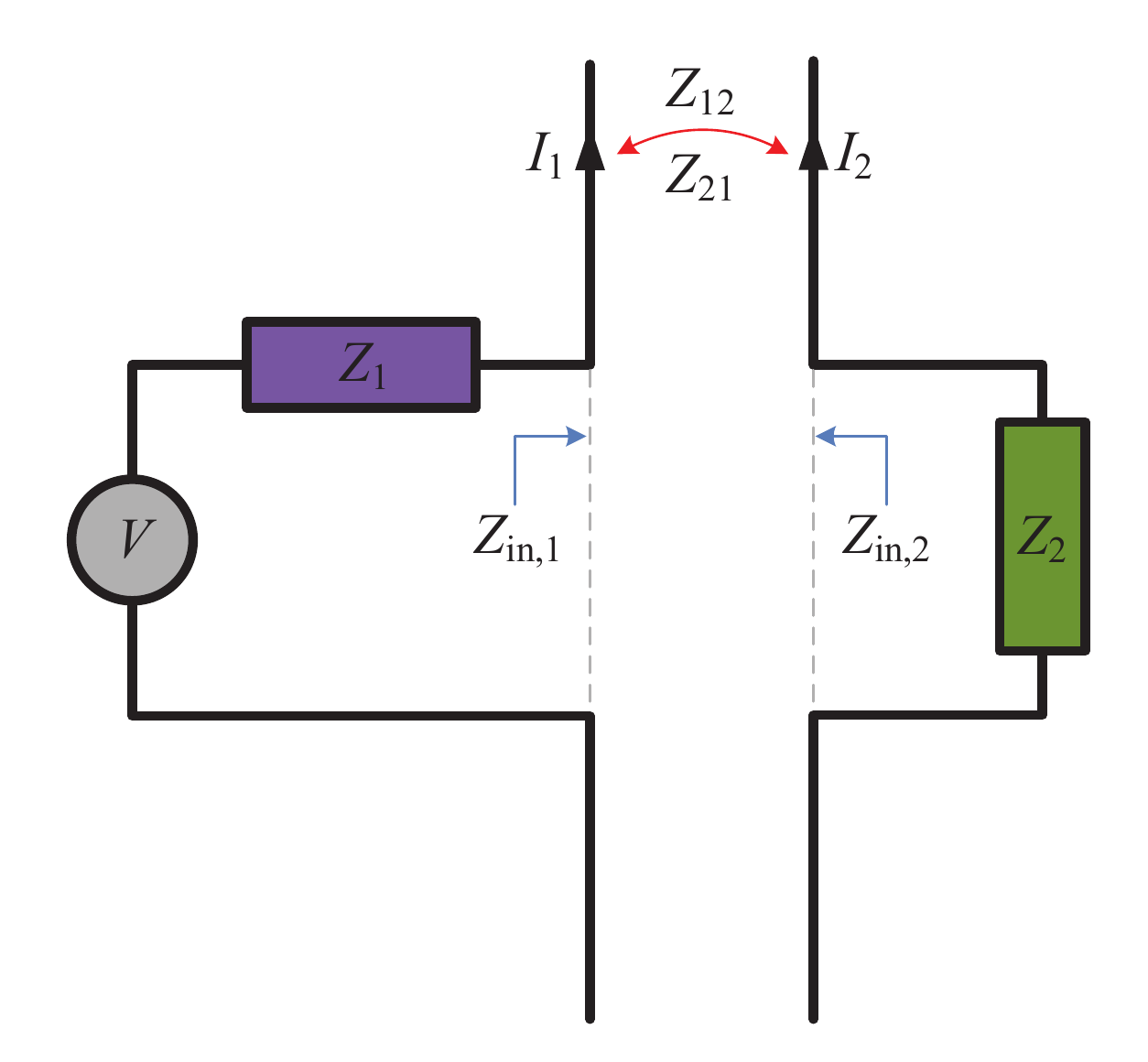}\label{sub00}}
\centering
\subfigure[]{\raisebox{10mm}
{\includegraphics[width=0.24\textwidth]{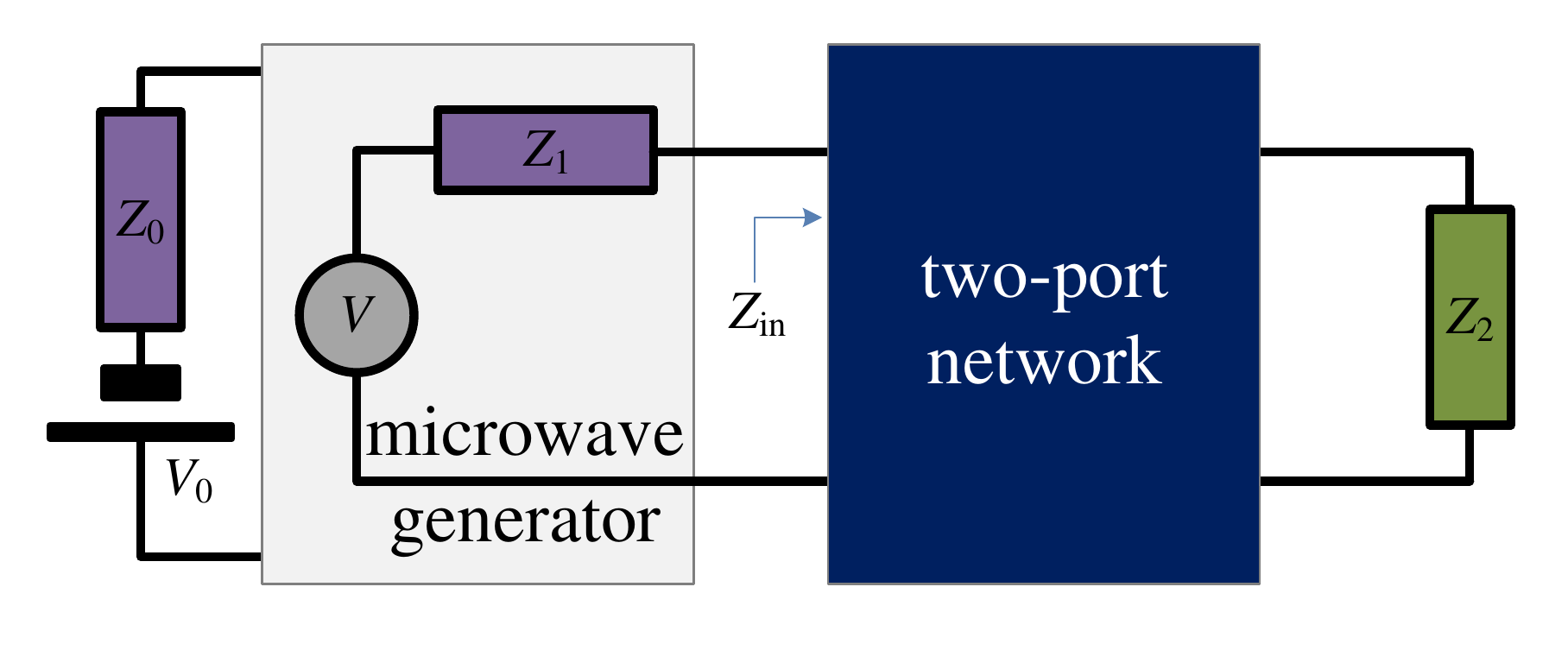}\label{sub01}}}
\caption{(a) Schematics of a conventional wireless transfer system. The wire dipoles can be replaced by two arbitrary antennas (e.g., loops). (b) An equivalent two-port representation.}
\label{fig0}
\end{figure}

This conventional paradigm is illustrated in Fig.~\ref{fig0}. In order to minimize power losses, the equivalent two-port network of the antennas and the free-space link between them should be reactive, which means that the antennas should be made of perfect conductors or other lossless materials, and the currents flowing on the antenna pair should form a non-radiating current distribution \cite{dewa,Lindell}, which is typically achieved operating in the very near field. In this way, radiation into free space is suppressed which, in the antenna engineering terminology \cite{Balanis}, is translated into nullifying the total radiation resistance of the antenna system. This nullifying is achieved by optimizing the geometry and positions of the two antennas \cite{no_rad1,no_rad2}. 

The overall efficiency of this power transfer is limited by the internal resistance of the generator $\Re(Z_1)$, which constitutes an inevitable factor of power loss. In this conventional paradigm for WPT, there are actually \emph{two} parasitic  internal resistances where energy is lost, because the generator $V$ in Fig.~\ref{sub01} is fed by a DC source or from the mains, which are also not ideal voltage sources and therefore contain some non-zero internal resistance $\Re(Z_0)$.

Two scenarios of power transfer optimization are usually considered. In one of them, the system is tuned to maximize efficiency, minimizing power absorption in all parts of the system except at the receiver load. To this end, all resistors except the load resistor at the receiving end must be made as small as possible, including  the internal resistance of the generator. In this scenario, the power efficiency can be large, however, the power output is not maximized, and the power extracted from the DC or 50/60 Hz supply and transferred to
the load is relatively low. In the second scenario, the system is tuned to maximize the power transfer into the receiver load, which requires conjugate matching of the load to the source. In this  case, half of the power is lost in the internal resistance of the source and the other half, in the ideal case, is delivered to the load. Therefore, even in the case of the ideal lossless coupling (e.g. via ideally conducting coils on a common lossless magnetic core of negligible electrical size) when the
overall WPT efficiency in its traditional definition \cite{book} is equal
100\% the overall efficiency of the system cannot exceed 50\%.

In both scenarios, maximization of delivered power requires mutual compensation of the reactive parts of the source, of the connecting two-port network, and of the load impedances, i.e., operating at resonance. Tuning the power transfer system to resonance as the load positions, environment, and impedance values vary requires complex control systems \cite{contr00,contr0,contr}. Besides the complexity, and, therefore, reduced reliability, the power consumed in these systems decreases the overall WPT efficiency.

In this paper, we introduce a different paradigm for wireless power transfer, in which the entire WPT system, including two antennas and the free-space link, forms a unified microwave generator. In this proposed scenario the load becomes an essential part of a microwave self-oscillating system. The optimal resonant regime is realized automatically as in a self-oscillating circuit. If we move or vary the receiving part of the system, the microwave frequency at which the self-oscillations are established automatically shifts, and no external tuning is actually needed. Since the intermediate step of converting DC or 50/60 Hz power into microwave oscillations is not needed, there is no parasitic resistance $Z_1$ at the transmitting end, meaning that one of the two parasitic loss resistances is effectively eliminated. This inherent resilience is rooted in the suitable balance between loss and gain characterizing the system, and it is consistent with recent work in the area of parity-time symmetric systems \cite{Fleury,teleport}.

\section{Self-Oscillating Wireless Power Transfer Systems}

The self-oscillating wireless power transfer concept that we introduce here can be understood after  considering the equivalent circuit of a conventional wireless transfer system, illustrated in Fig.~\ref{sub00}. A basic circuit analysis shows that the currents induced on the two antennas satisfy the linear system
\e
I_1(Z_1+Z_{\rm in,1})+I_2Z_{12}=V,\l{first0}
\f
\e
I_2(Z_2+Z_{\rm in,2})+I_1Z_{21}=0.\l{second0}
\f
Assuming that the mutual coupling is reciprocal, we can denote $Z_{12}=Z_{21}=-Z_{\rm m}$. Here the  minus sign corresponds to the definition of  the positive direction for currents $I_{1,2}$ as in Fig.~\ref{sub00}. The current at the load reads
\e
I_2={Z_{\rm m}V\over (Z_1+Z_{\rm in,1})(Z_2+Z_{\rm in,2})-Z_{\rm m}^2},\l{I}
\f
and the power delivered to the load is equal to $P=|I_2|^2{\rm Re}(Z_2)$. The delivered power is maximized when the imaginary part of the denominator is zero, which corresponds to the resonance of the circuit. Furthermore, the real part of the mutual impedance $Z_{\rm m}$ (mutual resistance) should be made as large as possible in order to minimize the real part of the denominator and increase the numerator. By increasing the real part of the mutual impedance, we compensate the radiation resistance of the two antennas (the real parts of $Z_{\rm in,1}$) and, accordingly, we minimize the parasitic radiation into surrounding space. In the ideal limiting case (not realizable except the case of an ideal transformer mentioned in the introduction), which corresponds to the non-radiating condition for the coupled antenna system, i.e., when ${\rm Re}(Z_{\rm in,1}Z_{\rm in,2}-Z_{\rm m}^2)\rightarrow 0$, all the power radiated by
the generator antenna 1 is received by antenna 2, and the
delivered power at resonance is limited only by the resistances
of the source and the load.

\begin{figure}[t]
\centering
\subfigure[]
{\includegraphics[width=0.25\textwidth]{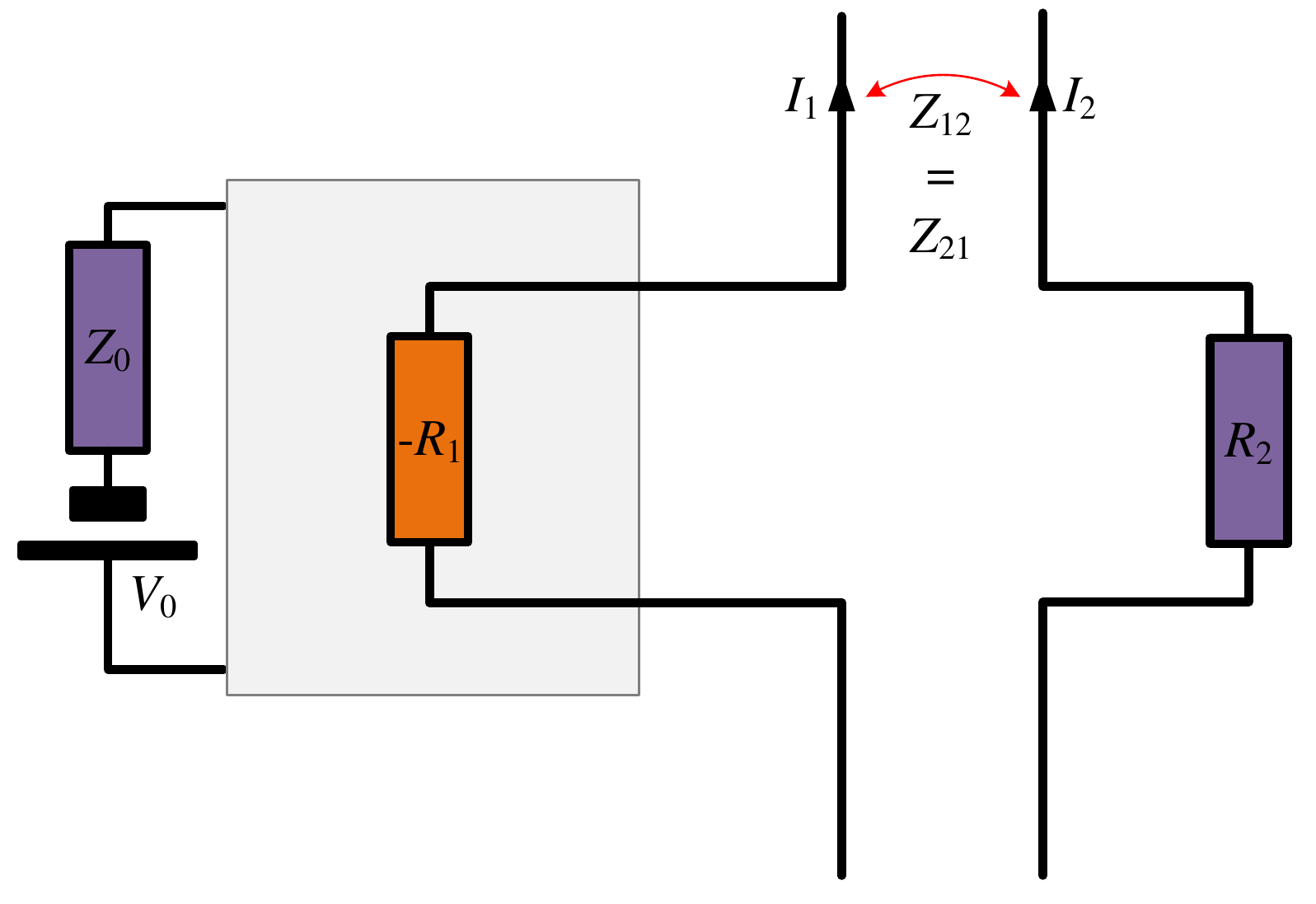}\label{sub4}}
\centering
\subfigure[]{\raisebox{7mm}
{\includegraphics[width=0.17\textwidth]{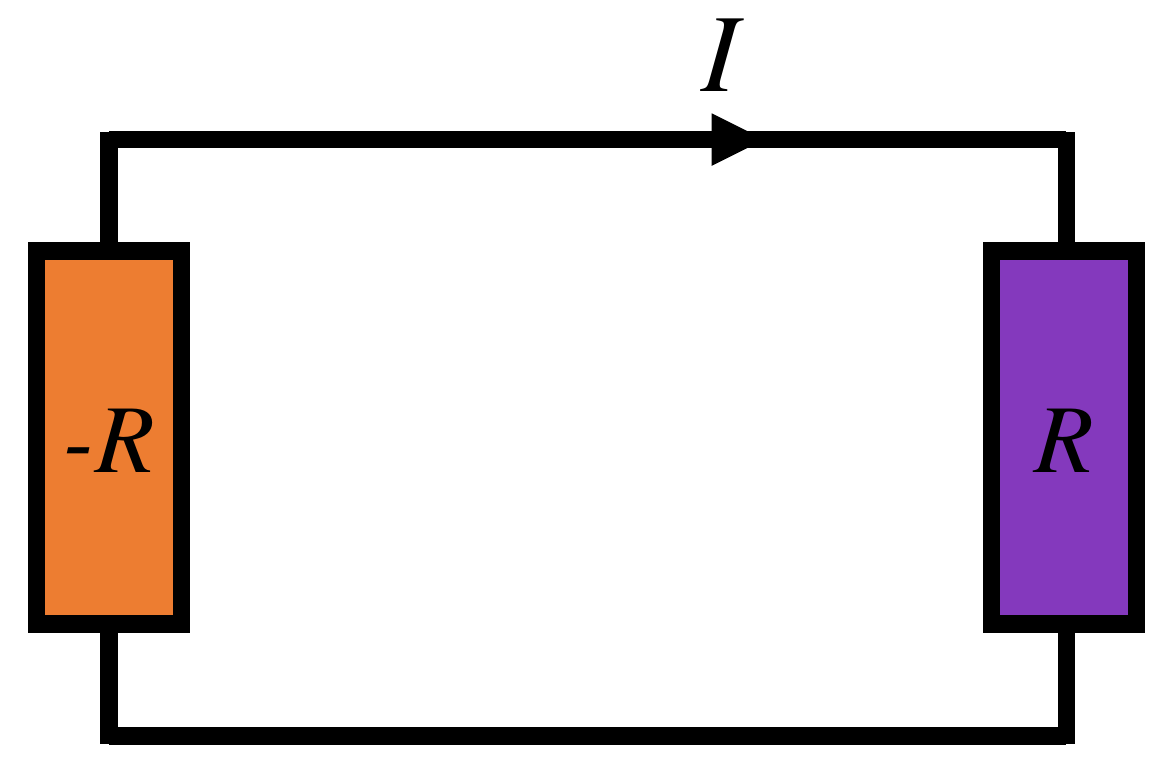}\label{sub5}}}
\caption{(a) Scheme of the proposed self-oscillating wireless power system based on a negative resistor. (b) Idealized equivalent scheme in the self-oscillating regime.}
\label{fig3}
\end{figure}

We propose instead a wireless power delivery system that forms a self-oscillatory system in which microwave power is directly \emph{generated} at the load, eliminating the limitation due to the source internal resistance $Z_1$. Considering the same conceptual equivalent scheme in Fig.~\ref{sub01}, this regime can be realized if we assume that the internal resistance of the first circuit can be actually negative and that there is no voltage source, as illustrated in Fig.~\ref{sub4}. In this case, the denominator of Eq.~\r{I} can be made identically equal to zero, which implies that the current amplitude in the load is limited only by the non-linear properties of the negative resistor.

The necessary conditions for the existence of self-oscillations can be found equating the determinant of the system \r{first0}--\r{second0} to zero (with $V=0$).  Let us assume for simplicity that the two antennas are identical and denote their input impedances by
\e
Z_{\rm in,1}=Z_{\rm in,2}=R_{\rm A}+j X_{\rm A}.
\f
We write the results for real values of the two impedances $Z_{1}=-R_{1}$ and $Z_2=R_2$, $R_{1,2}>0$ (the reactances can be incorporated into $Z_{\rm in,1}$ and   $Z_{\rm in,2}$). Denoting $Z_{\rm m}=R_{\rm m}+jX_{\rm m}$ and equating the imaginary part of the determinant to zero, we get
\e 
X_{\rm m}=X_{\rm A}{2R_{\rm A}+R_2-R_1\over 2R_{\rm m}}.\label{eq:OptimalXm}
\f
Minimization of radiation into free space corresponds to the situation when the mutual resistance $R_{\rm m}$ nearly compensates the antenna radiation resistance, that is, $R_{\rm m}\approx -R_{\rm A}$ (note that $|R_{\rm m}|$ cannot be larger than $R_{\rm A}$ since the antennas are passive). In this case equation (\ref{eq:OptimalXm}) simplifies into
\e 
X_{\rm m}=-X_{\rm A}\left(1+{R_2-R_1\over 2R_{\rm A}}\right).
\f
Equating the real part of the determinant to zero gives
\e 
R_{\rm A}(R_2-R_1)-R_1R_2+X_{\rm A}^2{R_2-R_1\over R_{\rm A}}\left(1+{R_2-R_1\over 4R_{\rm A}}\right)=0.\l{larger}
\f
Obviously, there are solutions with $R_2>R_1$. At resonance, $X_{\rm A}=0$, the self-oscillation regime corresponds to
\e 
R_1=R_2{R_{\rm A}\over R_{\rm A}+R_2}.
\l{NLR}\f

In this system, there is only one power transformation from the primary power source (DC or mains) to microwave frequency oscillations. If the load practically needs to be fed by
a DC or 50/60 Hz voltage, the backward transformation from
the microwave range is, of course, necessary. However, it is not
a necessary condition for a WPT system. The only unavoidable parasitic loss is in the inevitably nonzero internal resistance of the battery or of the main power supply, used to create negative resistance. Moreover, there is no need to manually or electronically tune the system to resonance, optimizing delivered power. The self-oscillating system generates microwave power in the load at the frequency or frequencies dictated by Eq. (\ref{eq:NLR}). If the load or antenna reactance varies, due to changes in the environment or in the location of the load, the frequency of self-oscillations will be modified accordingly. For the purpose
of power transfer, such a change in the frequency is irrelevant:
the delivered power is defined by the oscillations amplitude,
which  is mainly determined by the nonlinearity of the negative
resistor and is weakly affected by the frequency change. In
the ideal case, the schematic in Fig.~\ref{sub4} can be simplified to the one in Fig.~\ref{sub5}. It is interesting to note that the system shown in Fig.~\ref{sub5} is indeed parity-time symmetric, and this property of resilience is similar to recent works focusing on parity-time symmetric systems \cite{Fleury,teleport}. 

Although the above explanation is  based on the use of negative-resistance  circuits, the same proposed principle can be realized using other  self-oscillating circuits. For example, consider a generator formed by a microwave amplifier with an appropriate positive feedback network. In this  alternative scenario the wireless link can be a part of the feedback circuit of a generator, which creates microwave radiation directly where the power is needed. In other words, we include the near-field wireless link as an integral part of the microwave amplifier, and generate the microwave oscillations directly at the receiver. Depending on the operational frequency, this feedback channel can be for instance in the form of a waveguide or a coaxial cable or a microwave transmission line. Conceptually, to convert such a conventional generator into a wireless power delivery system, one can let a part of the feedback signal propagate in space, and position the power-receiving object into the field of the feedback wave.

\begin{figure}[t]
\centering
\includegraphics[width=0.35\textwidth]{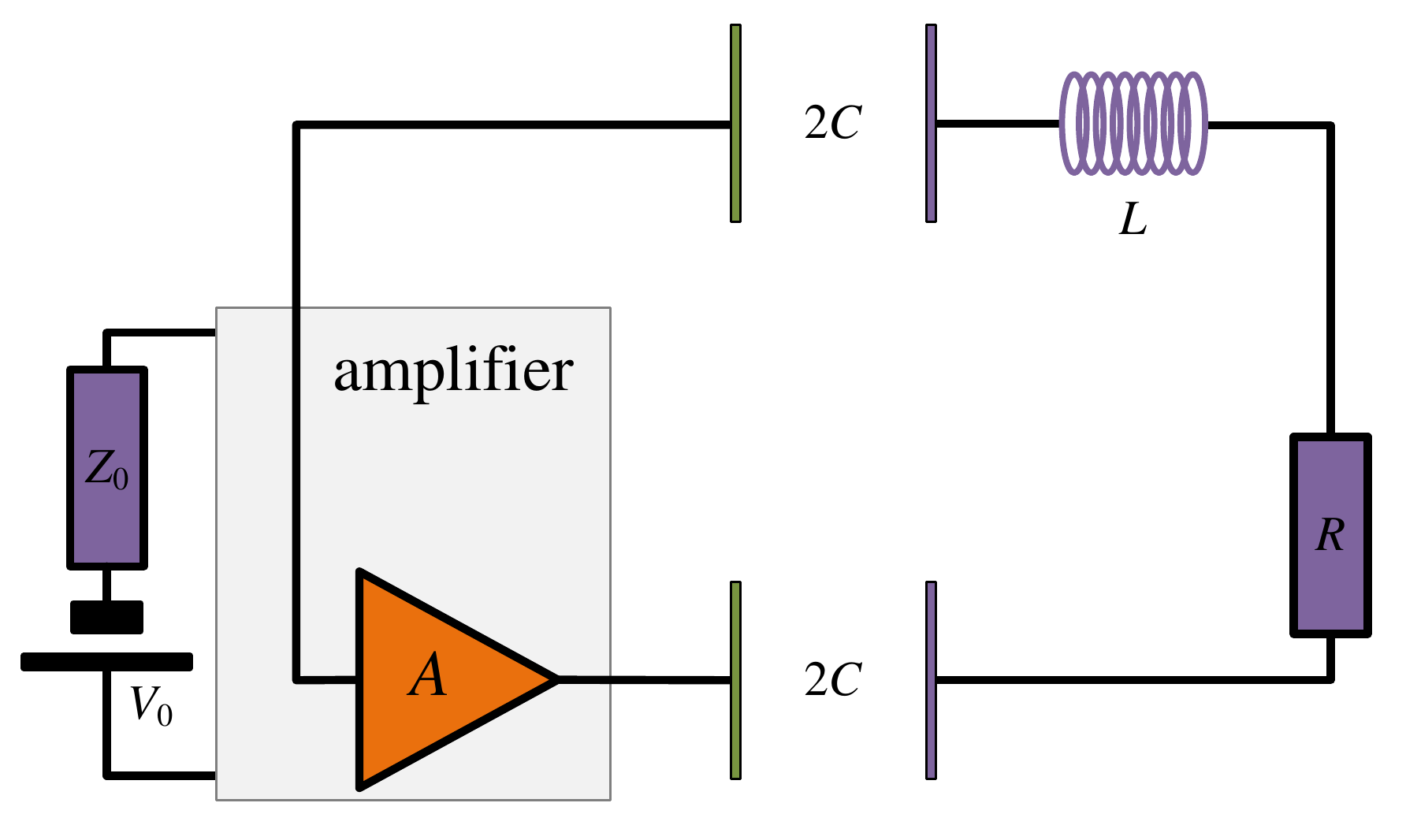}
\caption{A schematic of a self-oscillating  feedback power delivery system: A microwave amplifier with an over-the-air positive feedback system.}
\label{fig_new}
\end{figure}

As a particular example, consider a radio-frequency generator, where the positive feedback signal goes through a metal wire. Let us cut this feedback  wire and connect the two ends to two metal plates, positioning them on the same plane one next to the other. Since the impedance between the two plates is large, the feedback loop is effectively disconnected, and there are no oscillations. Thus, in this regime, very little energy is consumed from the power source that feeds the amplifier on stand-by. To deliver energy to a  resistive load, we equip the receiving device with other two metal plates of an appropriate size. If we now position this receiver on top of the two metal plates of the feedback circuit, at the frequency where the receiver plates, receiver connecting wires, and the feedback plates are in series resonance, the feedback loop closes and the self-oscillating system generates oscillations. Nearly all generated power is dissipated in the receiver only, if we use a high-efficiency amplifier.

This version of the proposed concept is illustrated in Fig.~\ref{fig_new}. Note that the only parasitic resistances are the loss resistances of the amplifier and the internal resistance of the primary power source which feeds the amplifier. The  internal resistance of the microwave generator [resistor ${\rm Re}(Z_1)$ in the previous discussion] has been eliminated. If the reactances or the load impedance change within a reasonable range, so that the series resonance of the  circuit falls within the frequency range where the amplifiers can operate, the same power will be delivered, even if the oscillation frequency changes.

\section{Numerical Demonstration of Wireless Power Generation}

\begin{figure}[t]
\includegraphics[width=0.5\textwidth]{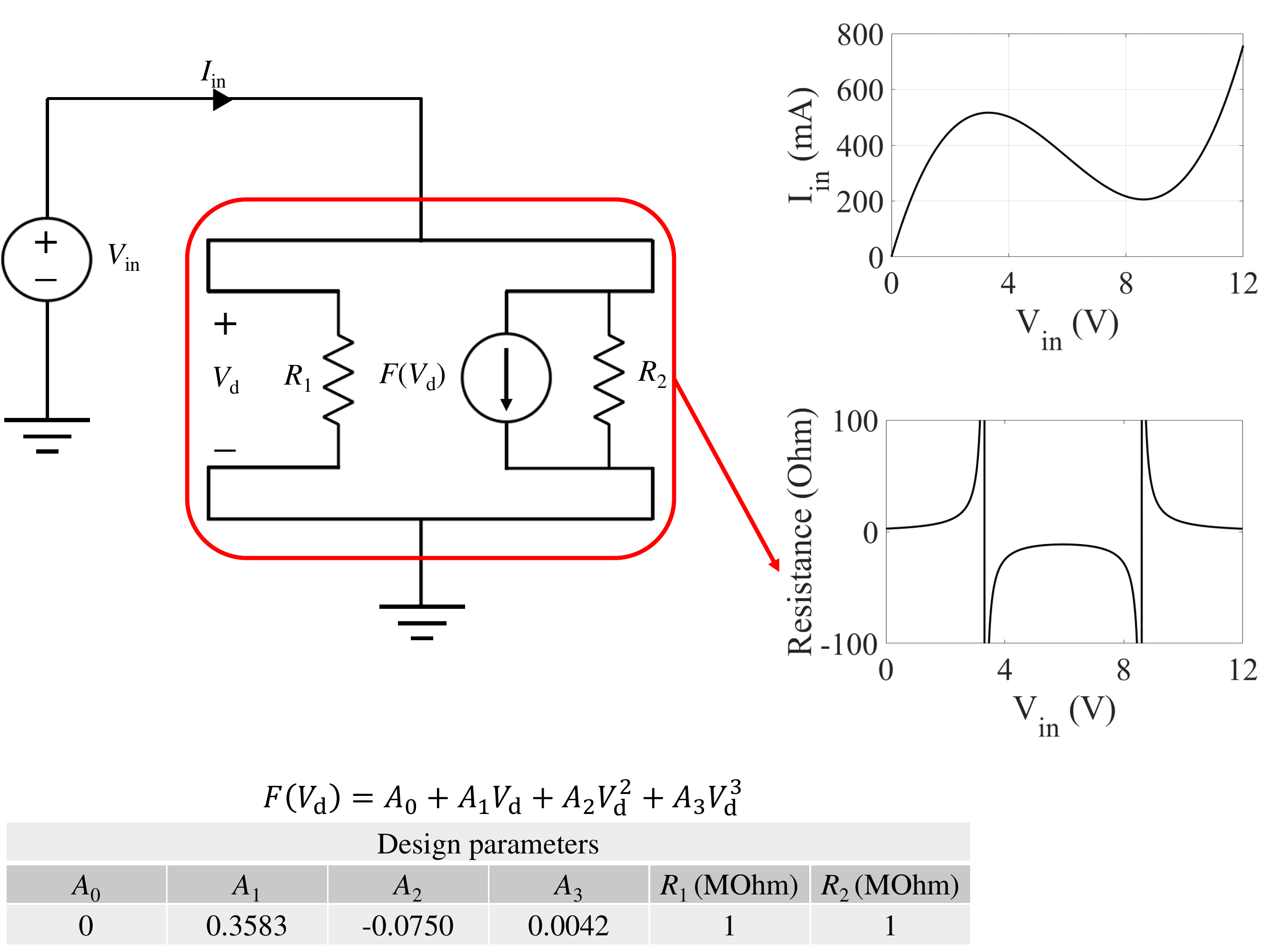}
\caption{Schematic of the circuit sub-element designed to provide non-linear differential negative resistance.}
\label{Figs-Y1}
\end{figure}

As a proof of concept demonstration of the potential of self-oscillating wireless power delivery, we consider a conceptually simple, one-dimensional numerical model. The active element is a non-linear negative resistor, realized as  a voltage-controlled current source  (see Fig.~\ref{Figs-Y1}.) The non-linear dependence of the current on the voltage is modeled by a third-order polynomial, with coefficients given in Fig.~\ref{Figs-Y1}. The same picture shows the corresponding volt-ampere curve and the associated differential resistance.

\begin{figure}[t]
\centering
\includegraphics[width=0.45\textwidth]{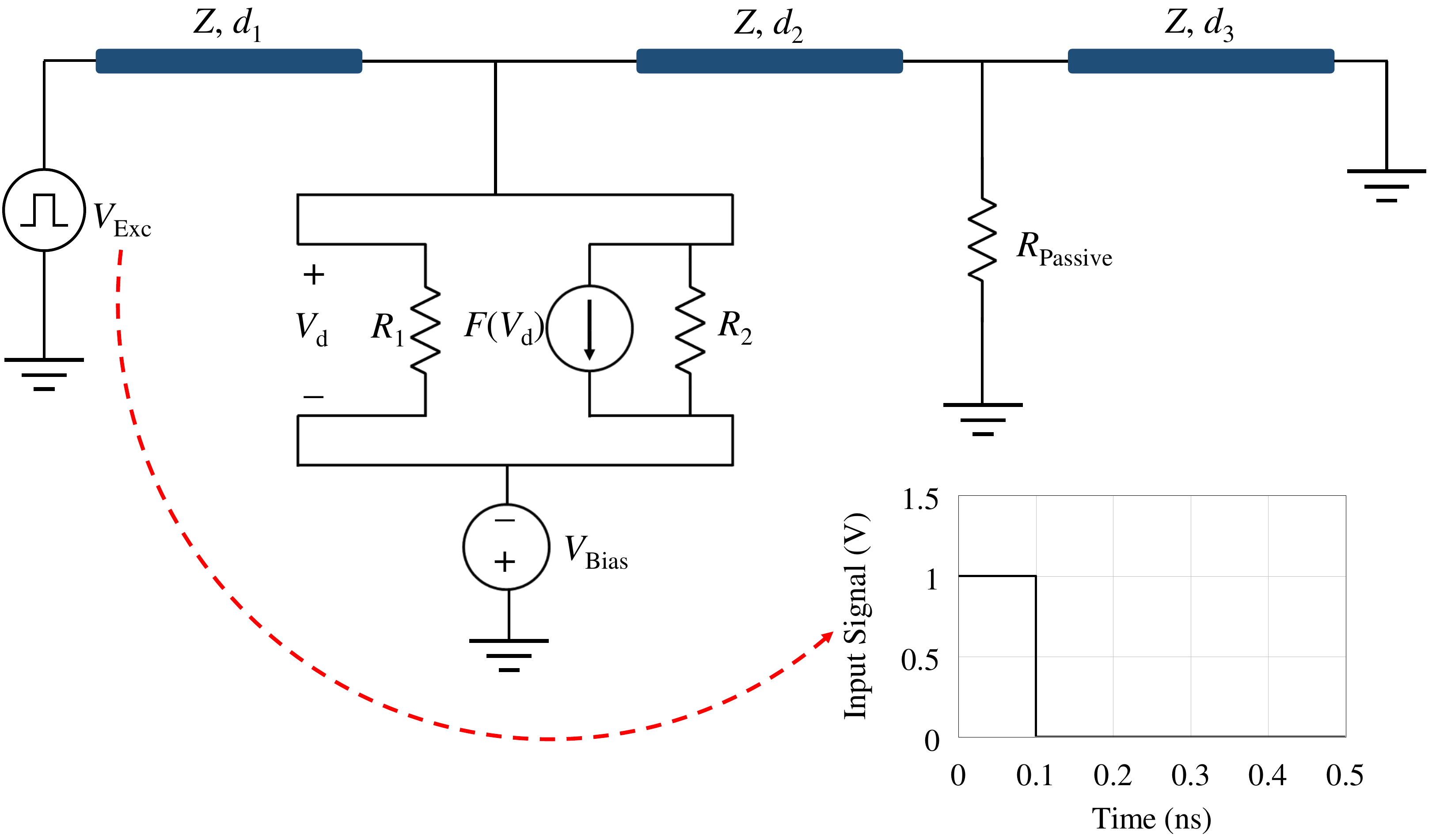}
\caption{Circuit model for the proof of principle concept.}
\label{Figs-Y2}
\end{figure}

Free-space propagation between the active element and the load $R_{\rm Passive}$, sustained by two antennas, is modeled by a transmission line of length $d_2$ and characteristic impedance $Z$  (see Fig.~\ref{Figs-Y2}). In this circuit, $V_{\rm Exc}$ pulse serves as the ignition that initiates self-oscillations in the circuit and dies out in less than $0.1$~ns (see subset in Fig.~\ref{Figs-Y2}). This initial push source models thermal noise in conventional negative-resistor oscillators, which is present at any frequency. In order to model the non-radiating mode of the two antennas of the wireless power transfer system we terminate the transmission line by short circuits at both ends. In this ideal case, there is no parasitic radiation into space. Obviously, this model is appropriate also for wireless power transfer between two objects inside a shielded room, or in the near-field of each other. Note that the characteristic impedance of the transmission line here can be arbitrary. As a proof of  concept example, here we assume the characteristic impedance of the transmission line to be $50$~Ohm. It should be noted that the proposed concept is perfectly applicable for more general scenarios where this impedance may be different depending on the particular realization. Note that for different scenarios the non-linear resistance is determined through equation (\ref{eq:NLR}).

\begin{figure}[t]
\centering
\includegraphics[width=0.45\textwidth]{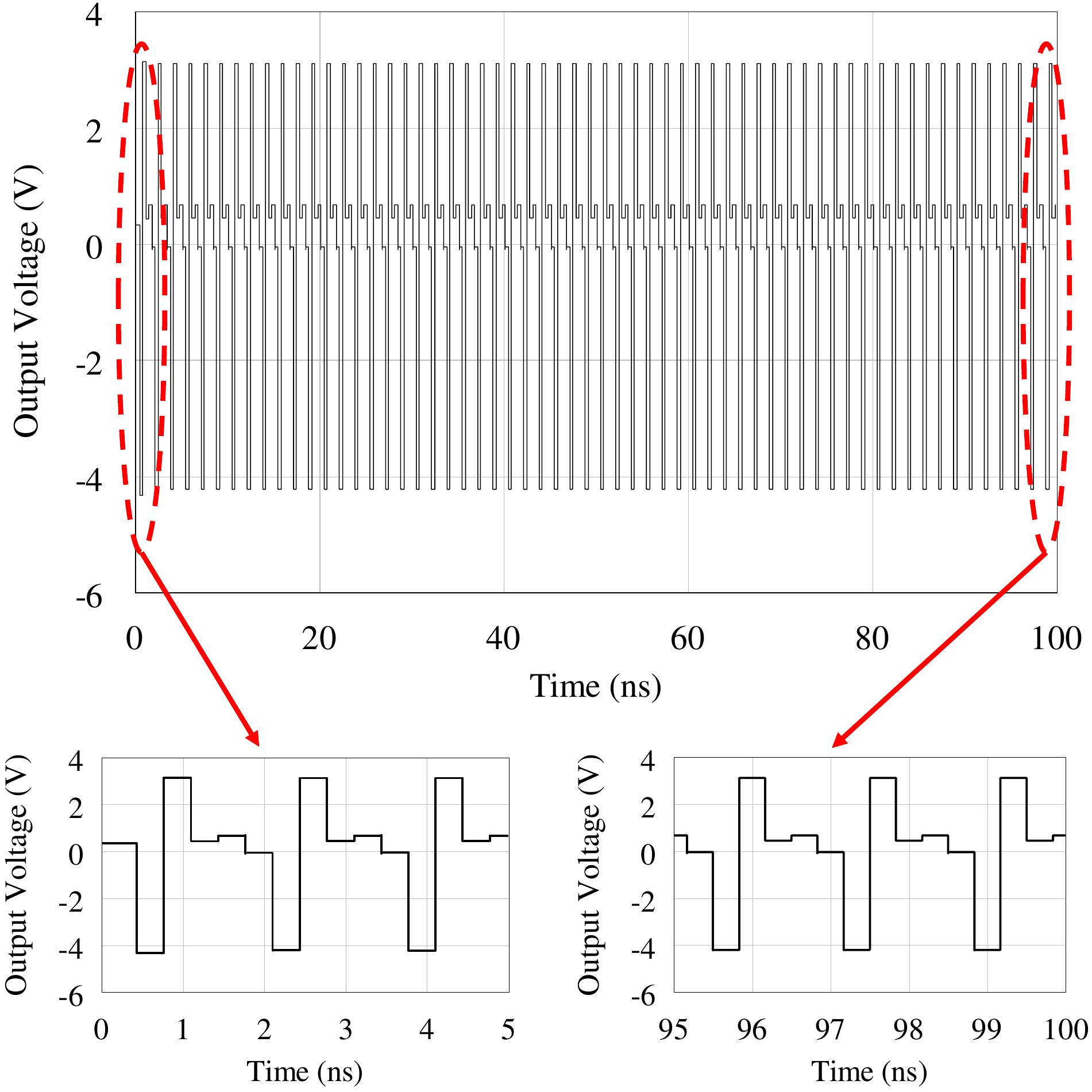}
\caption{Output voltage for the design shown in Fig.~\ref{Figs-Y2}. Design parameters: $Z=50$ Ohm, $d_1$=$d_2$=$d_3$=$50$ mm, $R_{\rm Passive}=50$ Ohm, and $V_{\rm Bias}=8.3$ V.}
\label{Figs-Y3}
\end{figure}
The bias voltage of the negative-resistor subcircuit ($V_{\rm Bias}$) should be chosen properly so that the non-linear subcircuit provides the required differential negative resistance and consequently the oscillations in the circuit can be initiated and continue after the push signal dies out. It can be seen from Fig.~\ref{Figs-Y1} that the differential resistance of the negative-resistor subcircuit depends on the applied bias voltage $V_{\rm in}$. Let us set the bias voltage of the subcircuit to $V_{\rm in}=V_{\rm Bias}=8.3$ V. As it can be seen from Fig.~\ref{Figs-Y1}, such a choice fixes the operational point of the subcircuit in the negative-resistance region. At this point, the differential resistance of the subcircuit is about $-54$ Ohm which, in the absolute value, is a bit higher than the resistance of the passive sheet ($R_{\rm Passive}=50$~Ohm in our example). Figure~\ref{Figs-Y3} shows the output voltage, that is, the voltage at the passive load. It is seen that the system is self-sustained after the push signal dies out, meaning that the negative resistor continues to deliver power to the passive load via the free-space propagation channel modeled by the transmission line.
\begin{figure}[t]
\centering
\includegraphics[width=0.45\textwidth]{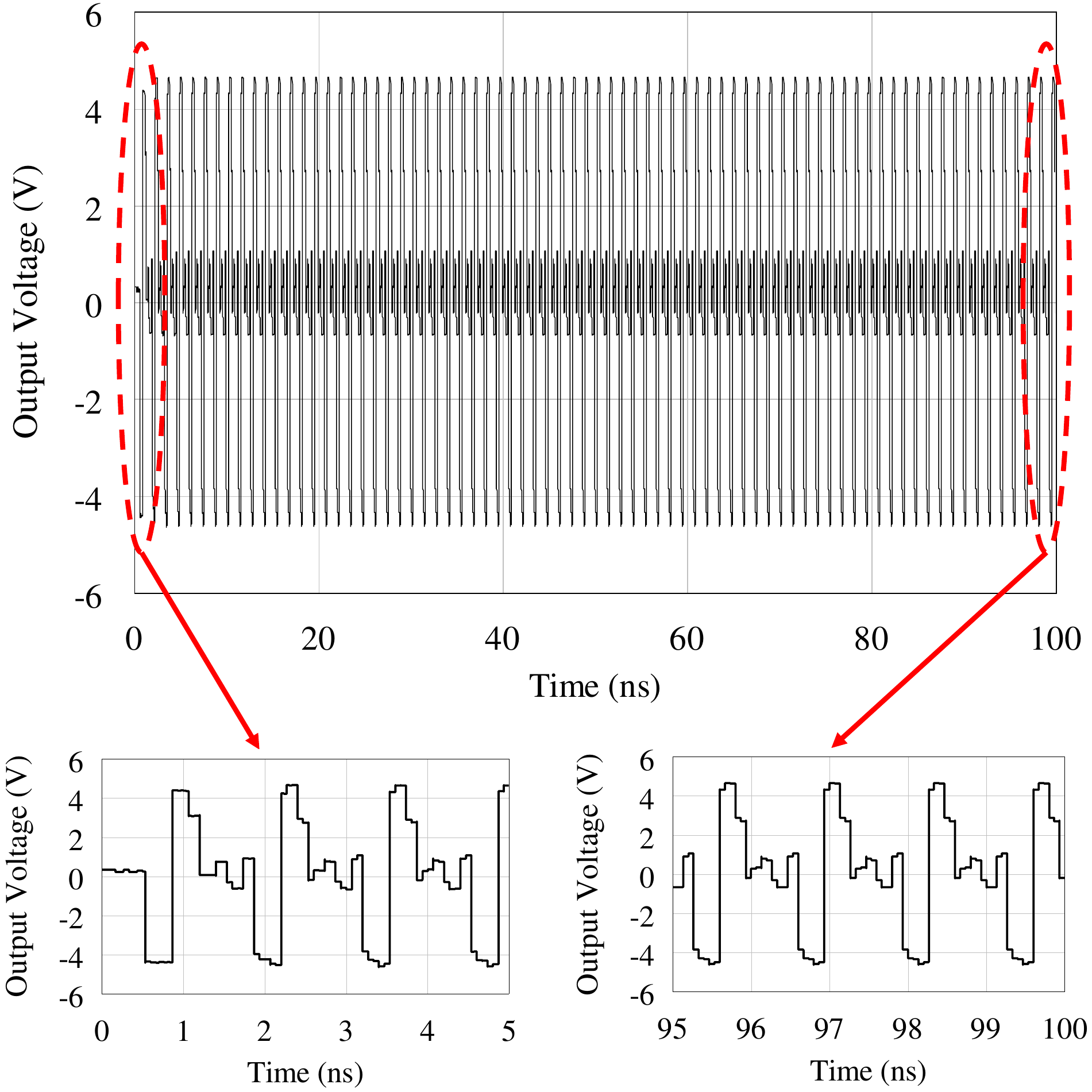}
\caption{Output voltage for the design shown in Fig.~\ref{Figs-Y2}. Design parameters: $Z=50$ Ohm, $d_1=d_3=50$ mm, $d_2=80$ mm, $R_{\rm Passive}=50$ Ohm, and $V_{\rm Bias}=8.3$ V.}
\label{Figs-Y6}
\end{figure}
\begin{figure}[t]
\centering
\includegraphics[width=0.49\textwidth]{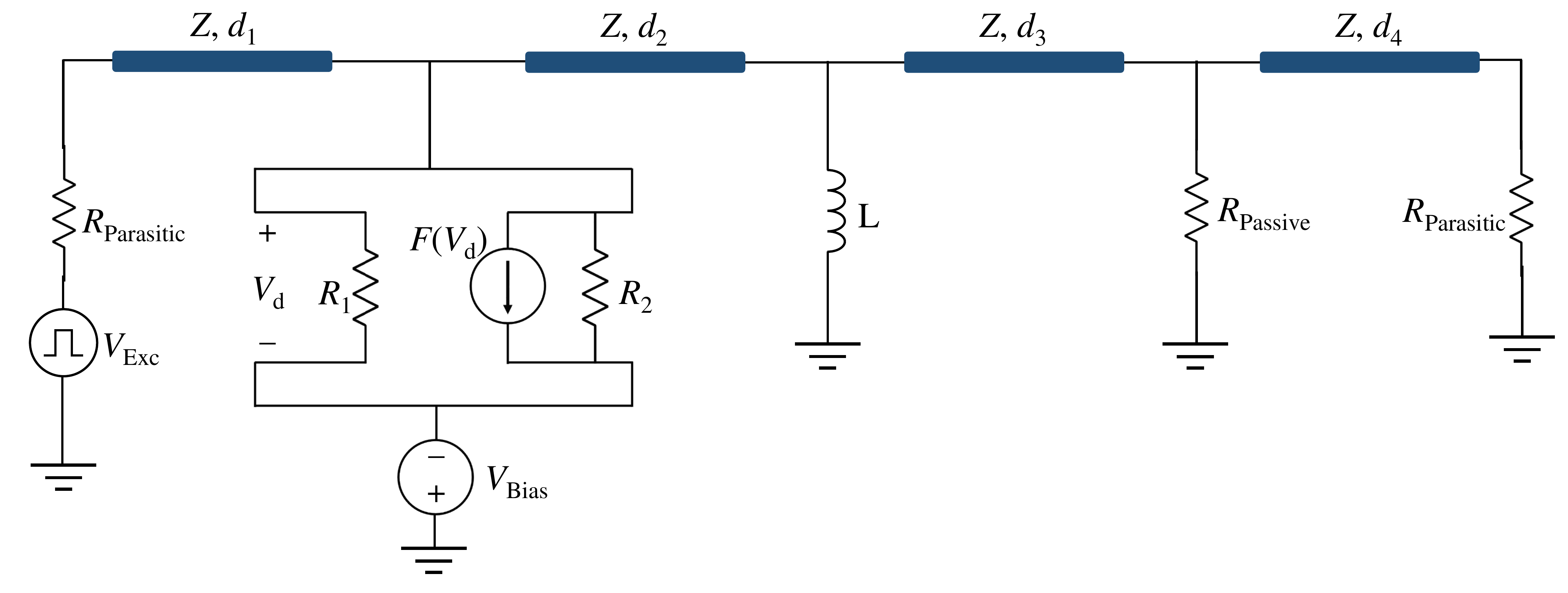}
\caption{Circuit model for the proposed concept, including a reflective obstacle (modeled by inductance $L$) and a parasitic loss resistor $R_{\rm Parasitic}$.}
\label{Figs-Y10}
\end{figure}
The result of Fig.~\ref{Figs-Y3} corresponds to the ideal scenario where the two antennas form an ideal transmission channel, without any parasitic radiation into space, and when there is no unwanted dissipation. Recently, a few studies have shown that the operation of the conventional WPT systems can be dramatically affected by perturbations in their environment \cite{no_rad1,Sawaya}. Thus, it is important to study the influence of the load positions, parasitic losses and energy leakage to free space. Let us first demonstrate that the proposed design is robust with respect to changes of the distance between the active element and the passive load. As an example, Fig.~\ref{Figs-Y6} shows the results for the case where the distance between the loads is increased. The results confirm that the oscillations continue in the circuit despite the change in distance, delivering power from the active element to the passive one. As expected, the frequency spectrum of the output signal changes, but the self-oscillation power delivery regime is very stable.
\begin{figure}[t]
\centering
\subfigure[]
{\includegraphics[width=0.45\textwidth]{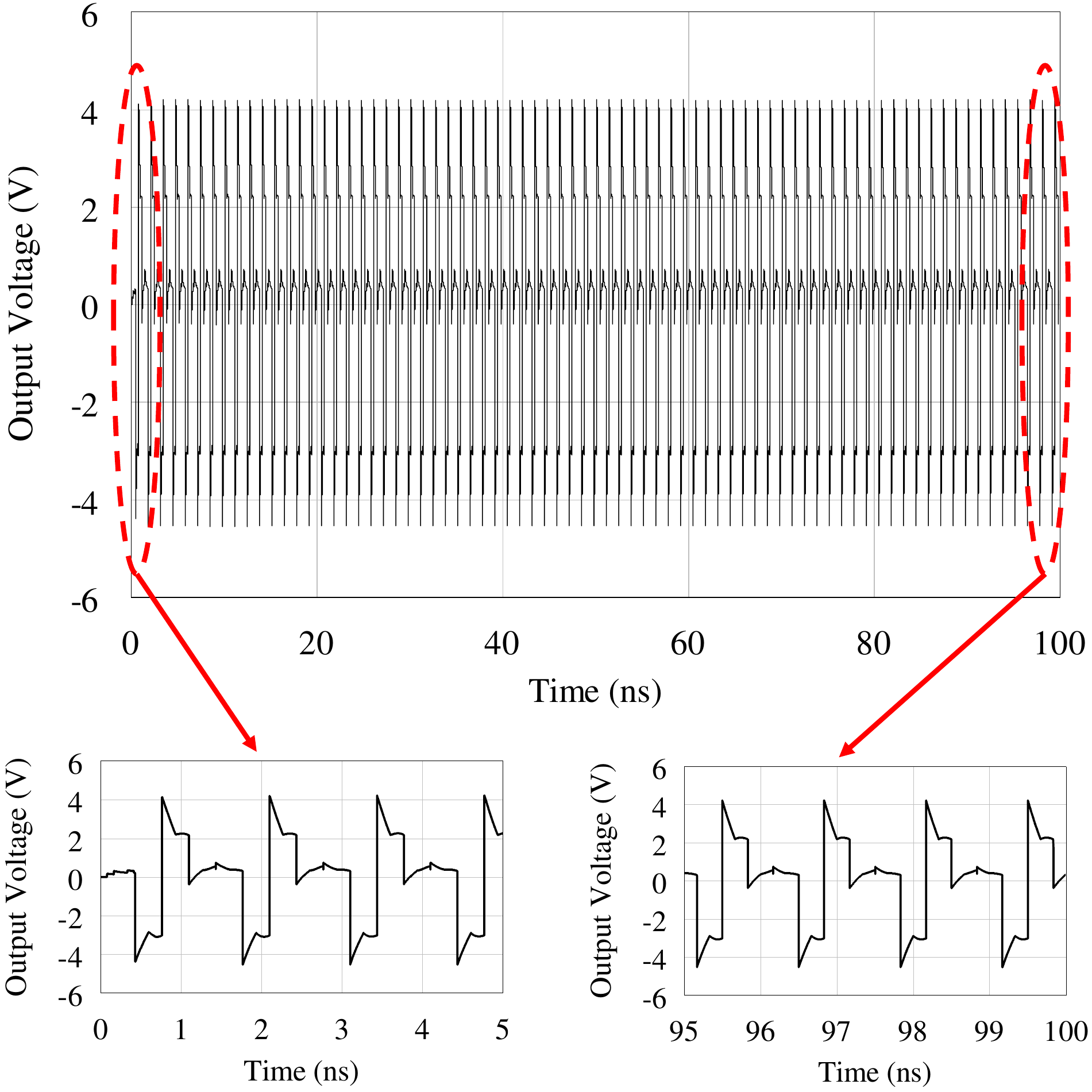}\label{Figs-Y8}}
\subfigure[]
{\includegraphics[width=0.45\textwidth]{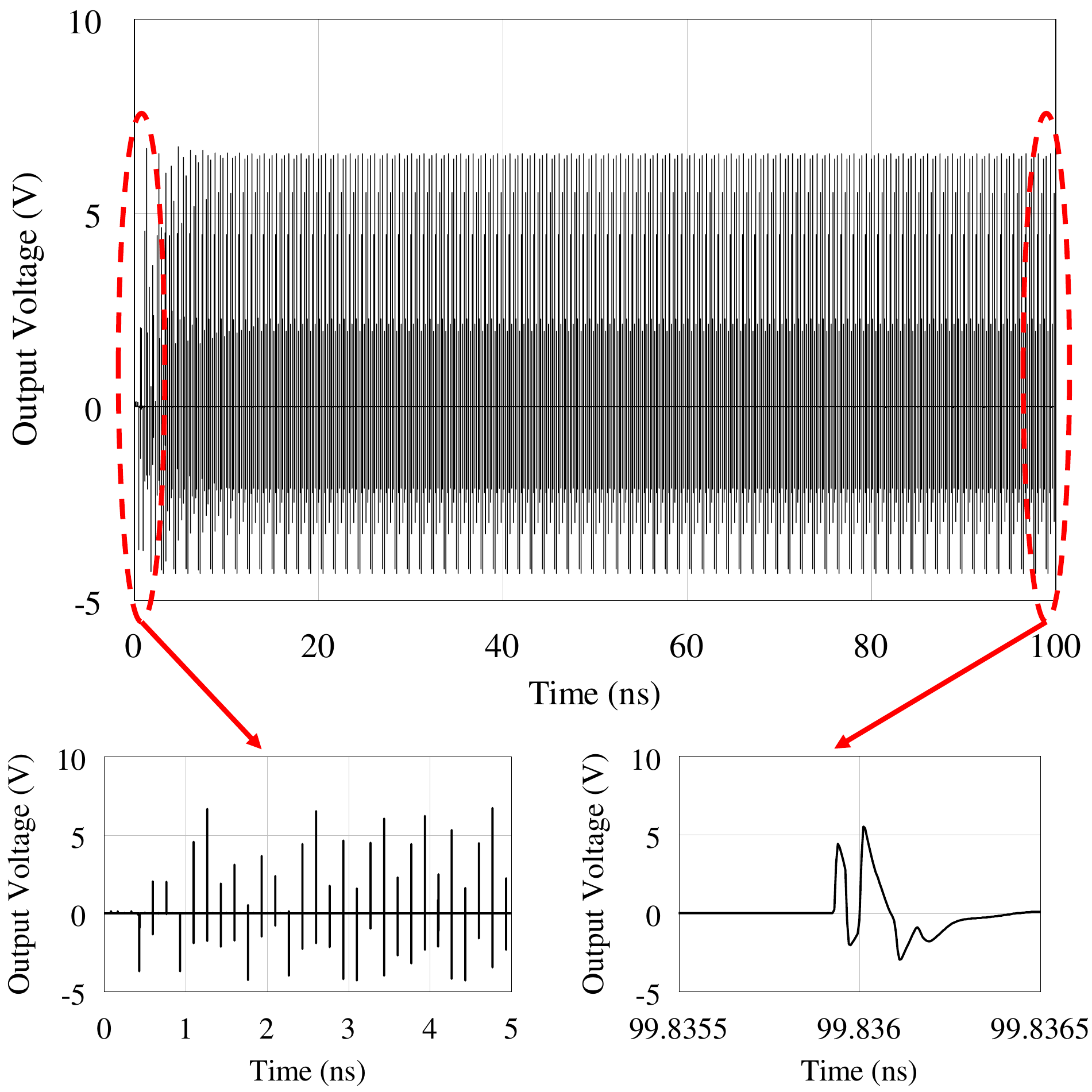}\label{Figs-Y9}}
\caption{Output voltage for the design shown in Fig.~\ref{Figs-Y10}. Design parameters: $Z=50$ Ohm, $d_1=d_4=50$ mm, $d_2=d_3=25$ mm, $R_{\rm Passive}=50$ Ohm, $V_{\rm Bias}=8.3$ V and (a) $L=10$ nH and (b) $L=1$ pH.}
\label{Figs-Y89}
\end{figure}

Next, we study the effects of presence of reflecting obstacles in between the source and the load. In the one-dimensional transmission-line model this corresponds to a reactive element, inductance $L$, inserted into the transmission-line path between the active element and the load (see Fig.~\ref{Figs-Y10}). In the plane-wave propagation scenario, such an inductive obstacle corresponds to a metal wire mesh or a metal screen perforated with electrically small holes, which strongly reflect the waves back to the source. In the same figure, we show parasitic resistors $R_{\rm Parasitic}$ that model parasitic radiation into surrounding space and dissipation losses. Figure~\ref{Figs-Y8} shows the results for the case in which an inductive element is located between the active and passive elements (here, $R_{\rm Parasitic}=0$). The structure is still able to deliver power from the active load to the passive one. Considering an even more extreme scenario, Fig.~\ref{Figs-Y9} shows the results for a highly reflective inductive element with  inductance  $L=1$~pH located in between the active and passive impedances. For plane waves in free-space, this inductive element is equivalent to an ideally conducting sheet perforated with circular holes of $0.85$~mm diameter and period $15$~mm, provided that the wavelength is considerably larger than the period. Results in Fig.~\ref{Figs-Y9} show that, although the inductive element reflects almost $99.99\%$ (numerical estimation at 10 GHz) of the impinging wave, the system will still establish and continue self-oscillations, self-tuning itself during the transient to the optimal condition for power delivery. Note that even in this extreme scenario, the amplitude of voltage oscillations in the load is not significantly affected.

\begin{figure}[t]
\centering
\includegraphics[width=0.45\textwidth]{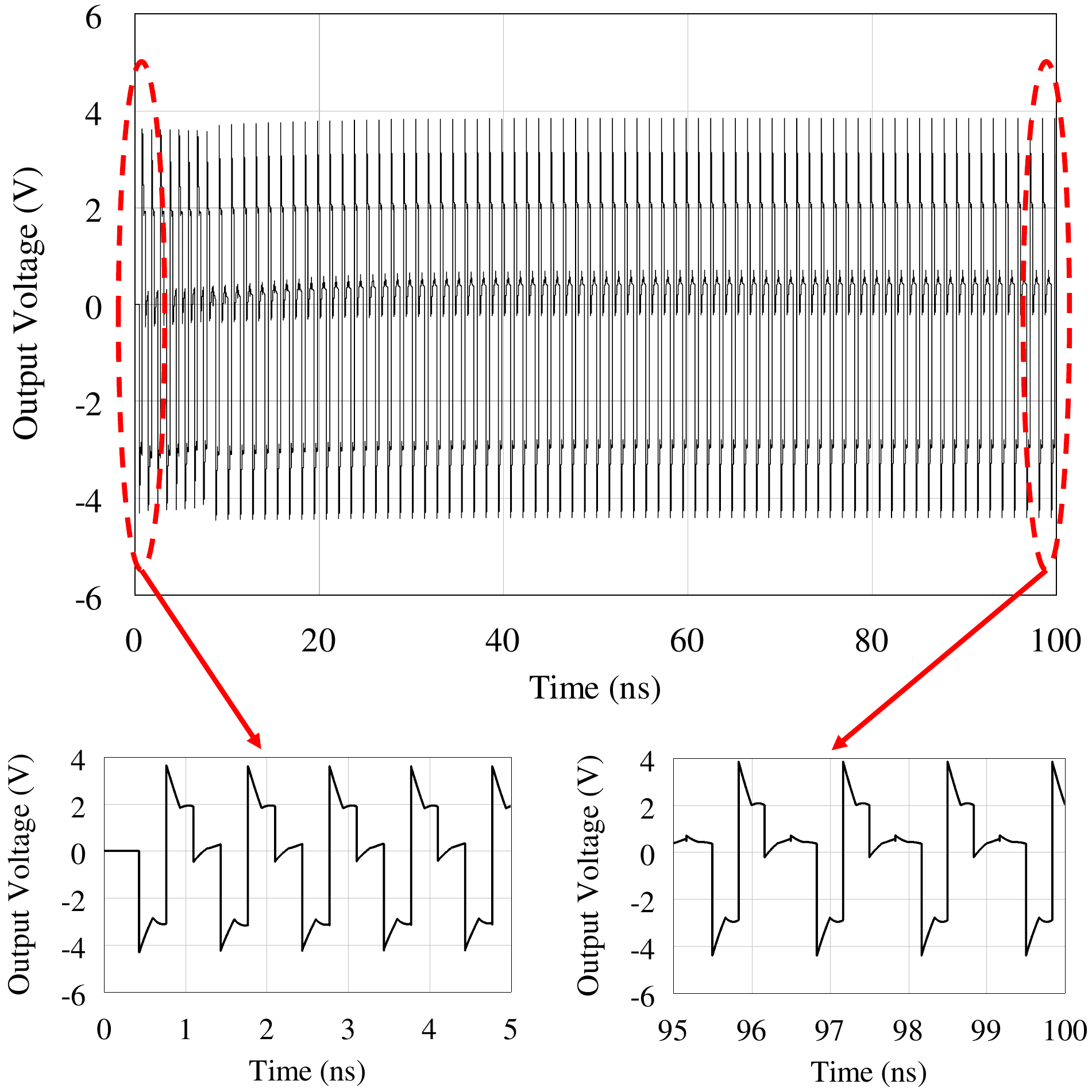}
\caption{Output voltage for the design shown in Fig.~\ref{Figs-Y10}. Design parameters: $Z=50$ Ohm, $d_1=d_3=50$ mm, $d_2=80$ mm, $R_{\rm Passive}=50$ Ohm, $L=0$, and $V_{\rm Bias}=8.3$ V. Parasitic loss resistor $R_{\rm Parasitic}=3$ Ohm.}
\label{Figs-Y11}
\end{figure}
Figures \ref{Figs-Y6} and \ref{Figs-Y89} demonstrate that if the radiation into the surrounding space is suppressed, the proposed self-oscillating power delivery system is extremely robust with respect to positions of the receiver and insensitive to the presence of obstacles between the transmitting and receiving antennas. This feature resembles the effect of ``teleportation'' of waves through nearly perfectly reflective walls, recently proposed in \cite{teleport}.

Finally, we study the detrimental effects of parasitic radiation and dissipation in the system. The results of Fig.~\ref{Figs-Y11} show the voltage oscillations in the load resistor when  the perfectly conducting boundaries at the two ends of the propagation channel model are replaced by resistors ($R_{\rm Parasitic}=3$~Ohm in this example). We see that  degradation of the voltage amplitude in the load is rather small. The presence of the parasitic loss resistor has stronger effect on the shape of the pulses (the frequency spectrum) than on the amplitude of the voltage oscillations in the load.

\section{Experimental Demonstration of a Self-oscillating Wireless Power Transfer System}

In the preceding sections we have shown that self-oscillating circuits (comprising microwave power generators) can be used as wireless power transfer devices if the positive feedback loop passes energy through space to the load. There exists a great variety of generators and most of them can be appropriately modified for the wireless transfer use. It is clear that different types of self-oscillating circuits will offer different advantages and the choice of a particular realization will depend on the application requirements. For example, the structure shown in Fig.~\ref{fig3} offers extremely good stability with respect to the position of the receiver and the presence of obstacles (as discussed above); however, it is difficult to realize high overall efficiency, because creation of negative resistance usually requires driving significant current through an Ohmic resistor (for example, in impedance inverter circuits). On the other hand, for the topology shown in Fig.~\ref{fig_new}, both the overall efficiency and stability with respect to the receiver position can be very good, but changes of the value of the load resistor will modify the self-oscillation regime. To make the study more complete and demonstrate the versatility of the proposed paradigm, for an experimental demonstration we have selected an alternative topology of self-oscillating systems which is immune to changes of the load impedance, while the delivered power somewhat changes when the transfer distance changes. 

\begin{figure}[h!]
	\centering
	\subfigure[]
	{\includegraphics[width=0.2\textwidth]{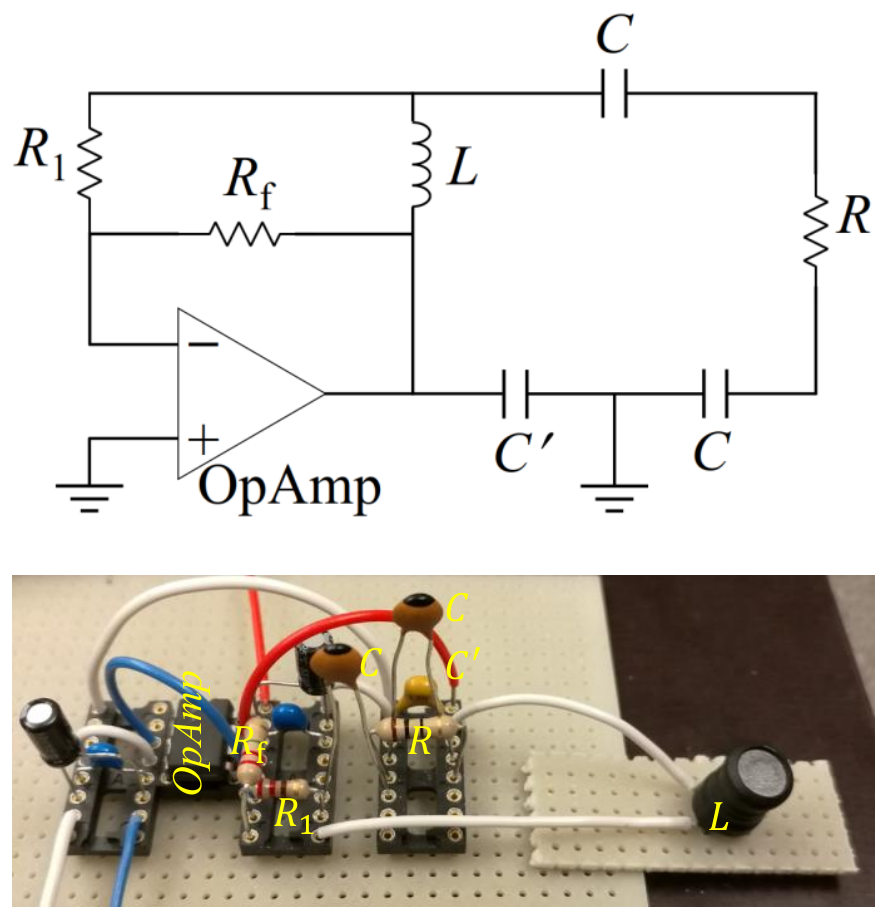}\label{fig11a}}
	\centering
	\subfigure[]
		{\includegraphics[width=0.28\textwidth]{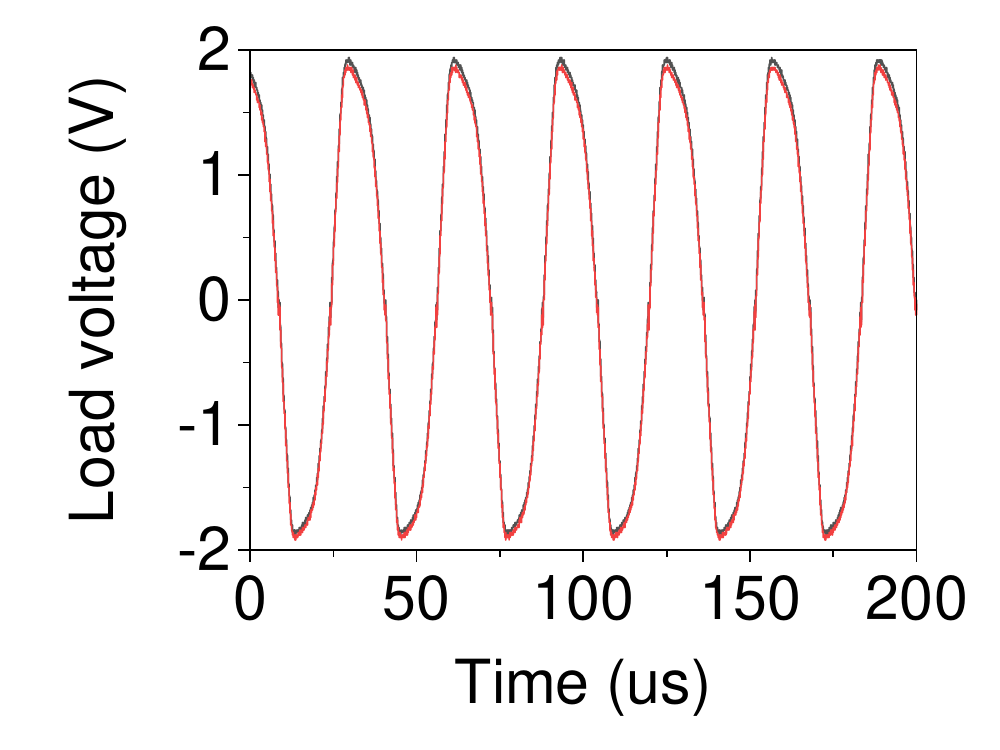}\label{fig11b}}
	\caption{(a) Schematic and experimental setup of a wireless power transfer system based on the Colpitts oscillator \cite{Basak,Gottlieb}. (b) Voltage oscillations on the two sides of the load resistor $R=1$~Ohm when the capacitance is $C=470$~pF.}
	\label{fig11}
\end{figure}

The circuit which we study is a modified basic oscillating circuit based on an operational amplifier (we use an LM7171 amplifier from Texas Instruments). The schematic and the photo of the analyzed, constructed, and tested circuit are shown in Fig.~\ref{fig11a}. In this design, the two capacitors $C$ model the wireless channel (when they are replaced by metal plate pairs) through which the power is transferred to the load. For a simple demonstration, we used lumped elements with values $R_{\rm f}=2.2$~MOhm, $R_1=220$~Ohm, $L=100$~mH, and $C'=470$~nF. In the following, we study the oscillation period and the power consumption by the load with different load resistances $R$ and capacitances $C$ (modeling possibly varying loads and wireless transfer distances). The black and red curves in Fig.~\ref{fig11b} (values close to each other) show the voltages at the two ends of the load when $R=1$~Ohm and $C=470$~pF (modeling power transfer between two metal plates with dimensions of 50 cm$\times$50 cm and separation  of $4.71$~mm). We can see that quasi-sinusoidal oscillations are created at the load and the oscillation period is $T\cong32$~$\mu$s. Moreover, knowing the voltage drop $\Delta V$ across the load, we can calculate the averaged power consumption by the load by calculating $\frac{1}{RT}\int_0^T \Delta V^2 dt$, which gives 2.4~mW. 
However, the overall efficiency is low, as the total power supplied from the DC source equals 375~mW. There are two main reasons for this high power consumption in the circuit. One is the chosen operational amplifier, which consumes a significant amount of energy. The other reason is the chosen topology of the self-oscillating circuit in which the capacitor $C'$ directly connects the output of the amplifier to the ground. As the output is an oscillating wave, there is a non-negligible AC current passing through $C'$ to the ground, leading to high power loss in the output resistance of the amplifier. We note that the power consumption by the other lumped elements ($R_{\rm f}$, $R_1$, $L$ and $C$) is very low compared to the load resistance. Therefore, the overall efficiency of the WPT system based on self-oscillating circuit can be improved by using a better amplifier and alternative topologies of self-oscillating circuits.


\begin{figure}[h!]
	\centering
	\subfigure[]
	{\includegraphics[width=0.24\textwidth]{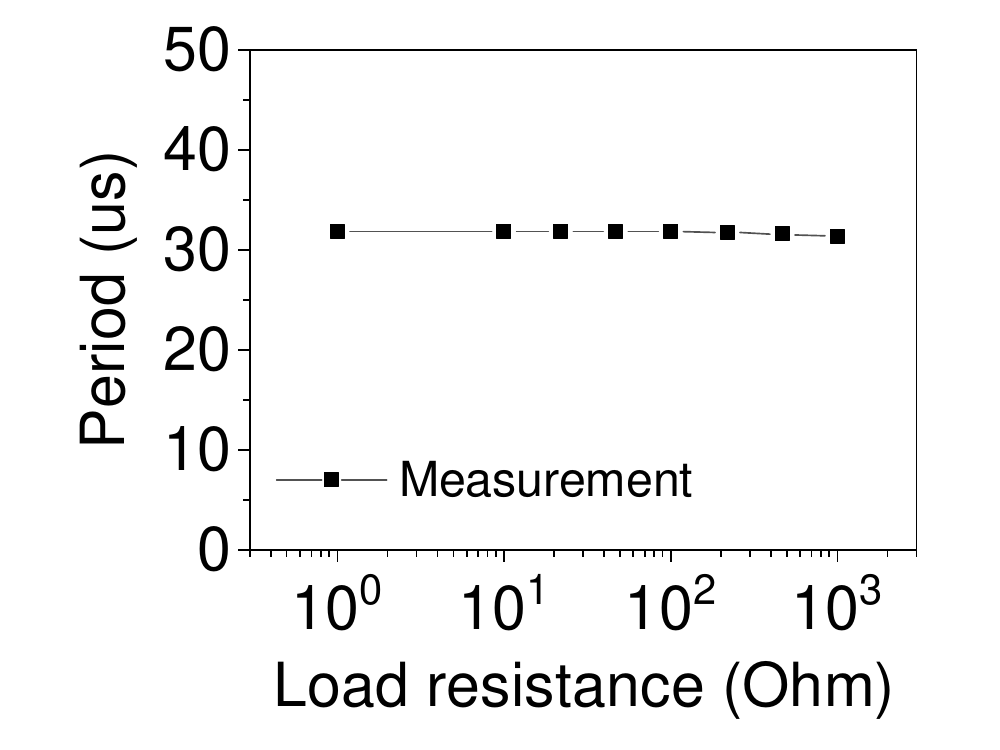}\label{fig12a}}
	\centering
	\subfigure[]
		{\includegraphics[width=0.24\textwidth]{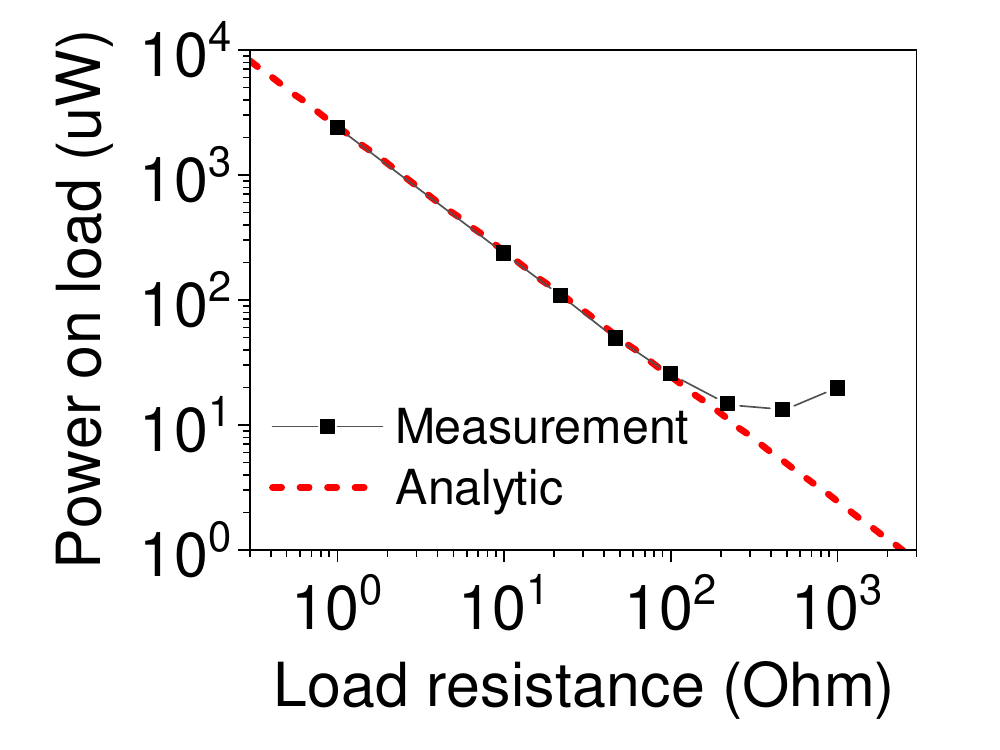}\label{fig12b}}
	\caption{(a) Oscillation period and (b) power consumption for different load resistances when $C=470$~pF.}
	\label{fig12}
\end{figure}

Next, we perform the same measurements for different load resistances while keeping $C=470$~pF (modeling a fixed distance to the load). The measured oscillation period and power consumption for different $R$ are shown in Fig.~\ref{fig12}. It is clear that the period exhibits remarkable stability with respect to $R$ which admits operation at a fixed frequency with a changing load. We see that the wireless system creates a virtual near-ideal voltage source at the load position, and the power consumption at the load decreases as $1/R$  when $R\leqslant 220$~Ohm simply because the oscillating voltage across $R$ is almost unchanged. When $R>220$~Ohm,  the electromotive force at the load increases. The red dashed curve in Fig.~\ref{fig12b} is the plot of $\frac{1}{2} \Delta V_{\rm max}^2/R$ with $\Delta V_{\rm max}=70$~mV, showing that the WPT system indeed realizes a nearly ideal voltage source at the load position, for load resistances smaller than $220$~Ohm.

\begin{figure}[h!]
	\centering
	\subfigure[]
	{\includegraphics[width=0.24\textwidth]{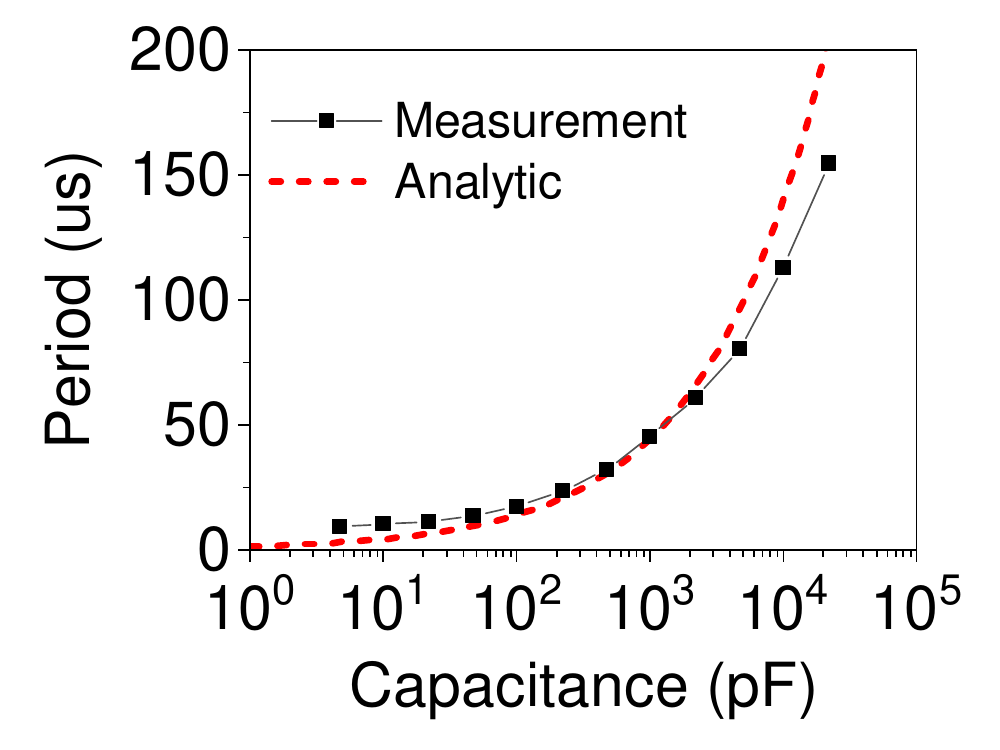}\label{fig13a}}
	\centering
	\subfigure[]
	{\includegraphics[width=0.24\textwidth]{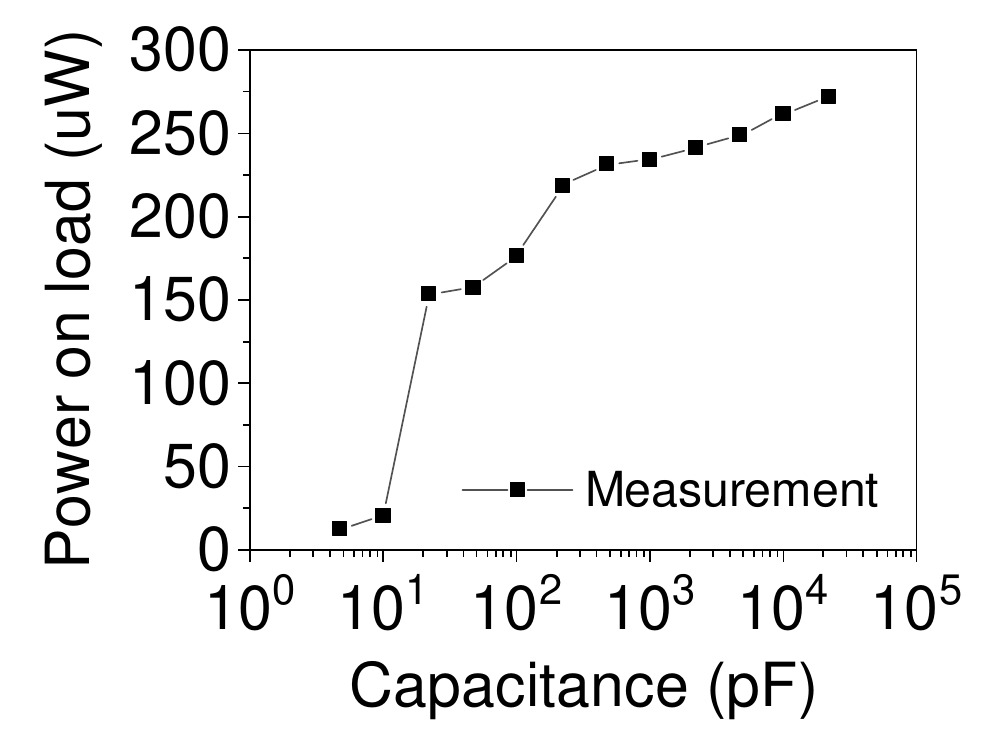}\label{fig13b}}
	\caption{(a) Oscillation period and (b) power consumption by the $R=10$~Ohm load for different capacitance values (different transfer distances).}
	\label{fig13}
\end{figure}

Finally, we study the power transfer when we vary the capacitance value $C$, corresponding to changes of the transfer distance in actual  WPT implementations. In this case, we keep the load resistance constant at $R=10$~Ohm and change the capacitance values from 4.7~pF to 22~nF (i.e., corresponding to change of the wireless transfer distance from 0.471~m to 0.1 mm for capacitors realized based on metal plates with area $50\times 50$~cm$^2$). The experimental results are shown in Fig.~\ref{fig13}. As we observe, the oscillation period increases when we increase $C$. Actually, the self-oscillation period can be obtained by analyzing the tank circuit, which gives $T=2\pi \sqrt{L C C'\/(C+2C')}$. This analytical period is plotted as the red dashed curve in Fig.~\ref{fig13a}, explaining the experimental results. On the other hand, the power delivered to the load increases as the capacitance gets larger (corresponding to decreasing distance to the load). It is due to the fact that a larger capacitance supports more charge accumulation inside the capacitor leading to a higher voltage drop across the load. The aforementioned results clearly show that this version of the proposed self-oscillating wireless transfer wireless systems is very efficient and extremely robust against changes of the receiver resistance while the delivered power changes with the transfer distance. We remind that the topology depicted on Fig.~\ref{fig_new} offers complementary properties, if those are preferable in the thought application.

\section{Conclusion}
In conclusion, we have presented an alternative paradigm of wireless power transfer, where the transformation of DC or mains energy into microwave oscillations is integrated with the wireless power delivery. Microwave energy is generated directly at the location where it is needed, although the primary power source is at a different location. Both conventional and proposed systems contain a microwave generator, which inevitably consumes some energy. But, the proposed solution contains \emph{only} a generator, while  conventional systems contain additional elements which also have their own parasitic losses (and the output impedance of the generator is inevitably one of these parasitic resistances). Realizing  power conversion from DC to oscillations and power transfer through space in one self-oscillating device offers possibilities for improvements of the overall efficiency, not available within conventional paradigms.  Moreover, this new approach  allows one to avoid the need of  active control feedback
loops that retune conventional WPT systems to the 
resonance when the environment of mutual locations of the
transmitting and receiving circuits change. As a result, the
robustness of the power transfer improves drastically. In the proposed designs, the system steady state  is self-established in an automatic way due to the non-linear properties of the active element (an amplifier or a negative resistor). The wireless link is a part of the feedback loop of a microwave self-oscillating circuit and the whole system is a single microwave generator.
In this paper we have numerically and experimentally
demonstrated three different versions of self-oscillating WPT
systems. We have compared the advantages and limitations of
the test versions.

As a figure of merit, the conventional definition of the efficiency of WPT systems (the ratio of the power delivered to the load to the power transmitted by the source antenna) loses its meaning completely, because microwave power is created directly at the load. For this reason, we define the efficiency as the ratio of the power delivered to the load to the power drawn from  the DC source (or mains). This overall system efficiency properly takes into account all power losses: in the active element, coupling elements, and due to parasitic radiation. In the realizations based on negative resistors (similar to PT-symmetric systems), there are parasitic losses in the active element, because in order to create negative resistance, bias current must flow through usual resistive elements. On the other hand, the feed-back scenario shown in Fig.~\ref{fig_new} is free from this limitation. With the use of high-efficiency amplifiers, there are no inevitable parasitic resistances except the internal resistance of the primary source (a DC battery, for example). Thus, we expect that the system efficiency of the WPT systems based on the proposed concept is inherently higher than their conventional counterparts. 

We believe that these results open new possibilities for the realization of simple, robust and effective wireless power transfer systems, and highlight a practically relevant application of self-oscillating circuits and balanced distributions of loss and gain for electromagnetics. 

During the review process of this manuscript, after our submission, Ref.~\cite{nature} was published, which independently from this work introduced a WPT scenario based on a combination of negative and positive resistors (similar to PT-symmetric structures), as a way to limit the sensitivity of WPT systems to variations in distance between source and receiver. As explained here, this is a special case of self-oscillating WPT systems, which we introduce here.

\section*{Acknowledgment}
This project has received partial funding from the European Union's Horizon 2020 research and innovation programme-Future Emerging Topics (FETOPEN) under grant agreement No 736876. Funding from NU ORAU Grant No. 20162031 and MES RK state-targeted program BR05236454 is also acknowledged.

\begin{IEEEbiography}[{\includegraphics[width=1in,height=1.25in,clip,keepaspectratio]{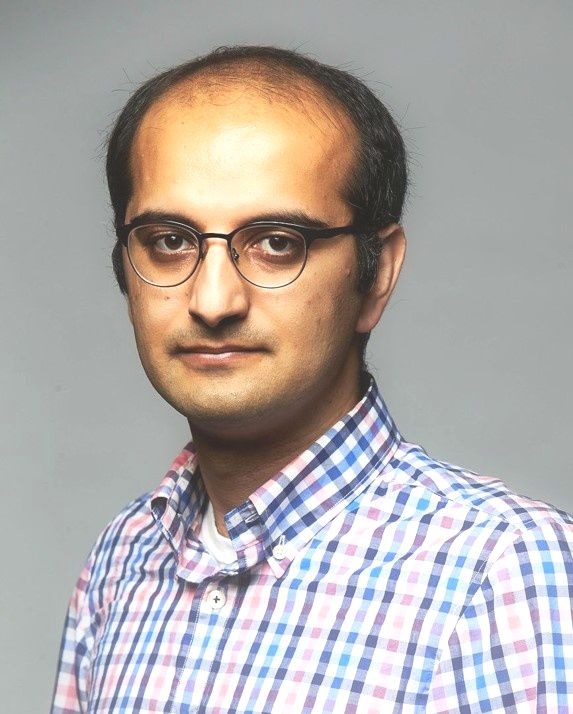}}]{Younes Ra'di} was born in Tabriz, Iran, in 1986. He received the B.Sc. and M.Sc. (with honors) degrees both in electrical engineering from the University of Tabriz, Tabriz, Iran, in 2009 and 2011, respectively. He received the Ph.D. degree (with distinction) in Electrical Engineering, Radio Science, from Aalto University, Espoo, Finland, in 2015, under the supervision of Prof. Sergei Tretyakov. During his doctoral studies his research was mainly focused on bi-anisotropic particles, optimal materials for strong light-matter interactions, and functional metasurfaces.

From Feb. 2016 to Sep. 2016, he was a postdoctoral research fellow in the group of Prof. Anthony Grbic at the University of Michigan, Ann Arbor, MI, USA, where he worked on Magnet-Free Nonreciprocal Metasurfaces and Tunable Metasurfaces. Since Oct. 2016, he has been a postdoctoral research fellow at the {\it Metamaterials \& Plasmonics Research Laboratory} led by Prof. Andrea Al\`u at The University of Texas at Austin, Austin, TX, USA.  

Over the last few years, Dr. Ra'di has received several research awards and grants including {\it Aalto ELEC Doctoral School Funding} (2013-2015), {\it HPY Research Foundation Grant} (2013), and {\it Nokia Foundation Grant} (two times in 2013 and 2014). He has received {\it Honorable Mention} at the Student Paper Competition of the {\it 2015 IEEE International Symposium on Antennas and Propagation and North American Radio Science Meeting}, Vancouver, Canada, and {\it 2nd Prize} at the Student paper competition of {\it The Ninth International Congress on Advanced Electromagnetic Materials in Microwaves and Optics-Metamaterials} in Oxford, UK, 2015. He has authored and co-authored more that 55 scientific contributions published in peer-reviewed journals and peer-reviewed conference proceedings. In particular, some of his papers have appeared in several high-impact journals including {\it Physical Review Letters} (Featured in Physics and Selected as Editor's Suggestion), {\it Physical Review X} (Featured in Physics and Selected as Editor's Suggestion), {\it ACS Photonics}, and {\it IEEE Transactions on Antennas and Propagation}. Some of his recent research works have been picked up by international media outlets.

His current research interests lie in the broad area of wave physics and engineering with emphasis on engineering characteristics of optical wave-matter interaction. In this context, he has made significant scientific contributions on a broad range of topics in theoretical and applied electromagnetics and optics, including artificial electromagnetic materials, bianisotropic media and inclusions, metasurfaces, thermal radiation, and translating fascinating light-wave phenomena from electromagnetics and optics to waves of different nature such as acoustic and elastic waves.

\end{IEEEbiography}

\begin{IEEEbiography}[{\includegraphics[width=1in,height=1.25in,clip,keepaspectratio]{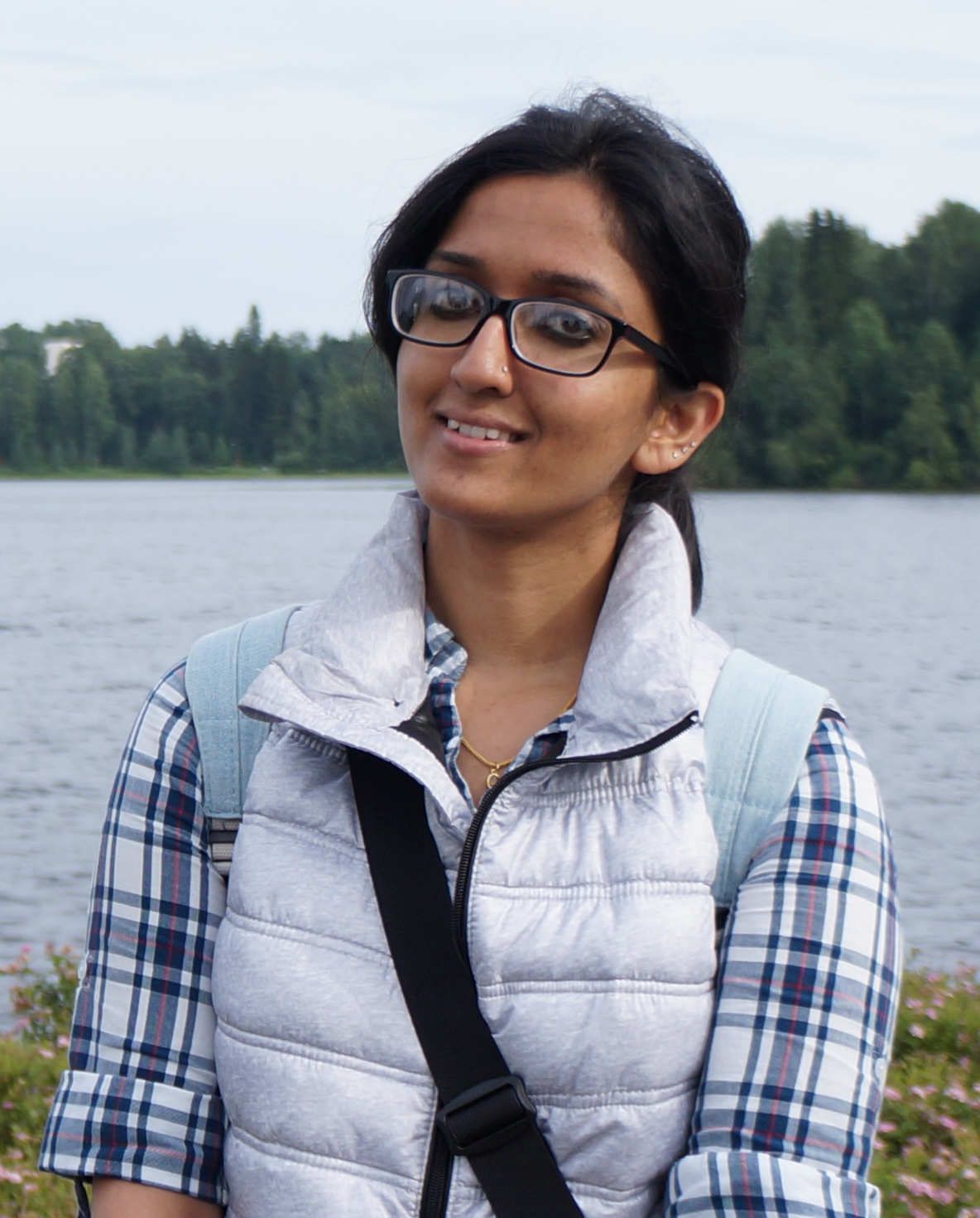}}]{Bhakti Chowkwale} received her Bachelor's of Engineering in Electronics \& Telecommunication from Savitribai Phule Pune University, India (Formerly known as University of Pune) in 2015. After graduation, she worked as an Engineer in Hella India Automotive Pvt Ltd. Currently she is working towards a Master of Science degree in Radio Engineering. Since 2017, she has been a Research Assistant for Professor Sergei Tretyakov in the field of Wireless Power Transfer. 
\end{IEEEbiography}

\begin{IEEEbiography}[{\includegraphics[width=1in,height=1.25in,clip,keepaspectratio]{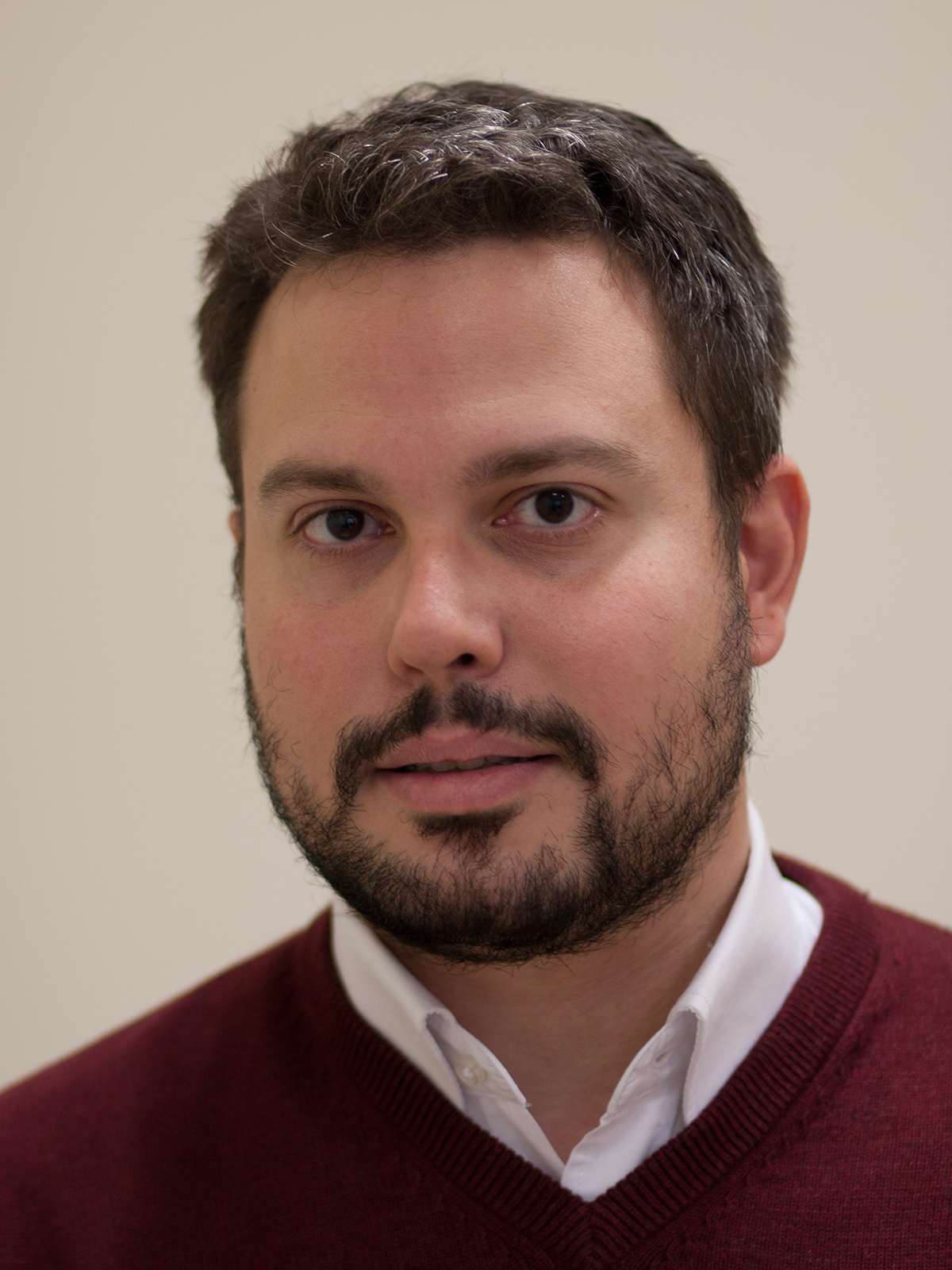}}]{Constantinos Valagiannopoulos} (M'12, SM'16) was born in Athens, Greece, in 1982. He received the Dipl.Eng. (Hons.) degree in Electrical Engineering from the National Technical University of Athens, Athens, Greece, in 2004, and the Ph.D. degree in Electromagnetic Theory in 2009. From 2010 to 2015, he was a Postdoctoral Researcher in the Department of Electronics and Nanoengineering, Aalto University, Espoo, Finland. During the academic year 2014-2015, he was with the Laboratory of Metamaterials and Plasmonics, Department of Electrical and Computer Engineering, University of Texas, Austin, TX, USA. From 2015, he is with Nazarbayev University, Kazakhstan, as an Assistant Professor with Department of Physics, School of Science and Technology. His research interests include artificial metamaterials, graphene, and their applications in photonic devices manipulating the light (absorbers, cloaks, lenses, etc). He has authored or coauthored more than 80 works in international refereed scientific journals and presented numerous reports in international scientific conferences.
Dr. Valagiannopoulos received the inaugural 2015 JOPT Research Excellence Award for his work: ``Perfect absorption in graphene multilayers''. He also received the International Chorafas Prize for the Best Doctoral Thesis in 2008, and the Academy of Finland Postdoctoral Grant for 2012-2015.
\end{IEEEbiography}

\begin{IEEEbiography}[{\includegraphics[width=1in,height=1.25in,clip,keepaspectratio]{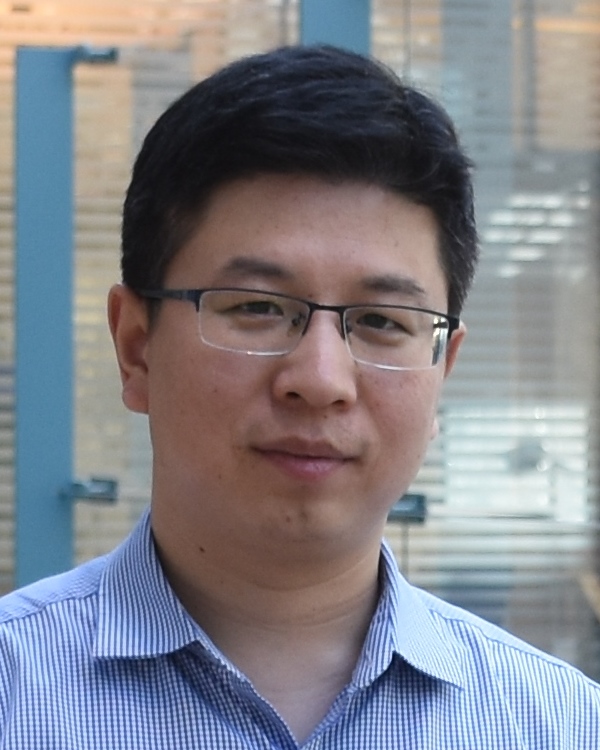}}]{Fu Liu} received the B.Sc. degree in applied physics from the China University of Mining and Technology, Xuzhou, China, in 2008, the M.Sc. degree in theoretical physics from the Beijing Normal University, Beijing, China, in 2011 and the Ph.D. degree in physics from the City University of Hong Kong, Hong Kong, China, in 2015.

From 2014 to 2016, he was a Research Fellow at the School of Physics and Astronomy, University of Birmingham, Birmingham, UK. Since 2017, he works as a Postdoctoral Researcher at the Department of Electronics and Nanoengineering, School of Electrical Engineering, Aalto University, Espoo, Finland.
\end{IEEEbiography}

\begin{IEEEbiography}[{\includegraphics[width=1in,height=1.25in,clip,keepaspectratio]{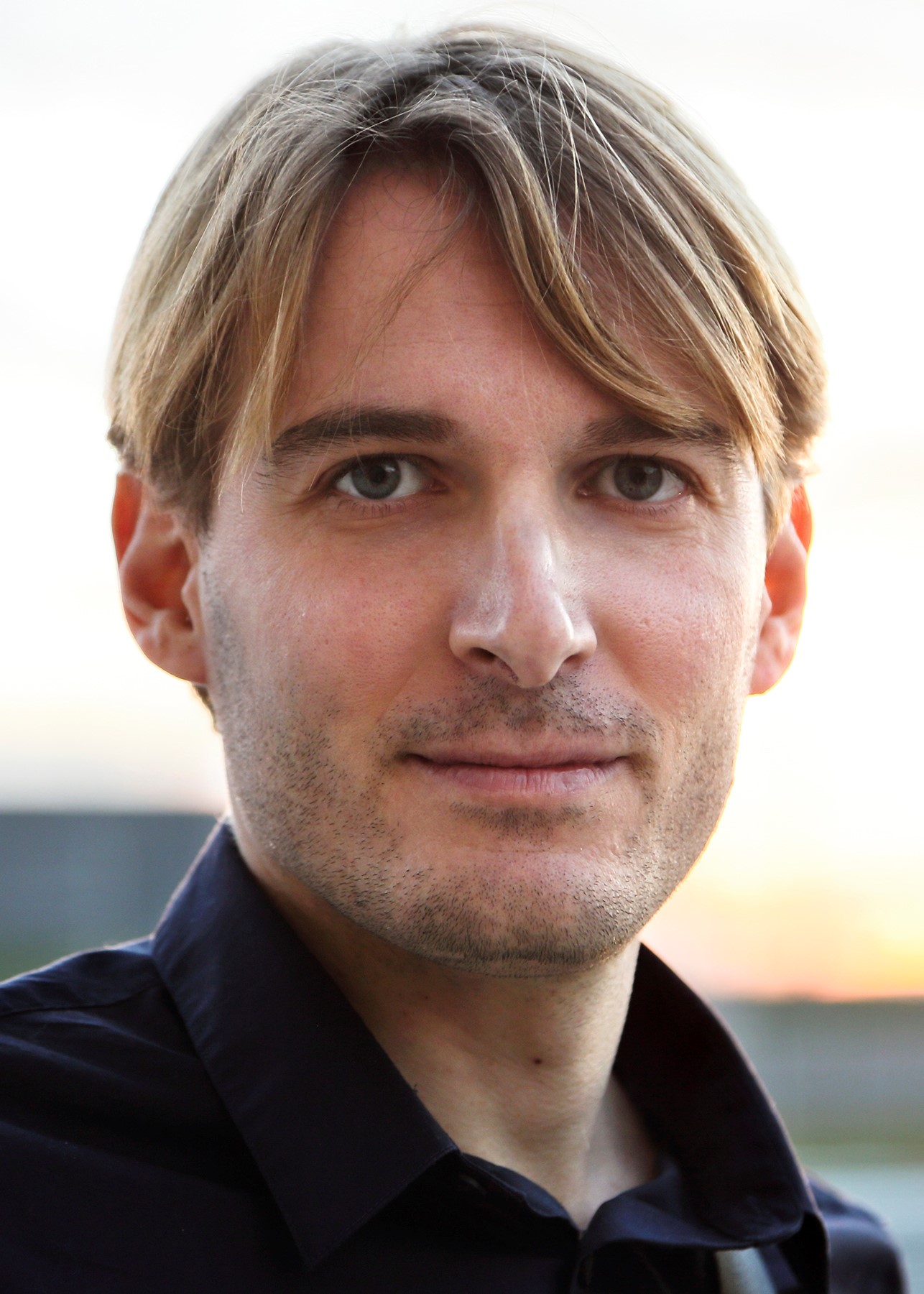}}]{Andrea Al\`u} (S'03, M'07, SM'12, F'14) is the Founding Director of the Photonics Initiative at the Advanced Science Research Center (ASRC) at the Graduate Center of the City University of New York (CUNY). He is also the Einstein Professor of Physics at the CUNY Graduate Center and Professor of Electrical Engineering at the City College of New York. He received the {\it Laurea}, MS and PhD degrees from the University of Roma Tre, Rome, Italy, respectively in 2001, 2003 and 2007. From 2002 to 2008, he has been periodically working at the University of Pennsylvania, Philadelphia, PA, where he has also developed significant parts of his PhD and postgraduate research. After spending one year as a postdoctoral research fellow at UPenn, in 2009 he joined the faculty of the University of Texas at Austin, where he was the Temple Foundation Endowed Professor until 2018. He is still an adjunct professor, senior research scientist, as well as a member of the {\it Wireless Networking and Communications Group}, at UT Austin.

He is the co-author of an edited book on optical antennas, over 400 journal papers and over 30 book chapters, with over 20,000 citations to date. He has organized and chaired various special sessions in international symposia and conferences, and was the technical program chair for the IEEE AP-S symposium in 2016, and for the international Metamaterials conference in 2014 and 2015. His current research interests span over a broad range of areas, including metamaterials and plasmonics, electromangetics, optics and nanophotonics, acoustics, scattering, nanocircuits and nanostructures, miniaturized antennas and nanoantennas, RF antennas and circuits.

Dr. Al\`u is currently on the Editorial Board of {\it Physical Review B}, {\it New Journal of Physics}, {\it Advanced Optical Materials}, {\it MDPI Materials}, {\it EPJ Applied Metamaterials} and {\it ISTE Metamaterials}. He served as Associate Editor for the {\it IEEE Antennas and Wireless Propagation Letters}, {\it Scientific Reports}, {\it Metamaterials}, {\it Advanced Electromagnetics}, and {\it Optics Express}. He has guest edited special issues for the {\it IEEE Journal of Selected Topics in Quantum Electronics}, {\it IEEE Antennas and Wireless Propagation Letters}, {\it Nanophotonics}, {\it Journal of Optics}, {\it Journal of the Optical Society of America B}, {\it Photonics and Nanostructures - Fundamentals and Applications}, {\it Optics Communications}, {\it Metamaterials}, and {\it Sensors} on a variety of topics involving metamaterials, plasmonics, optics and electromagnetic theory. 

Over the last few years, he has received several research awards, including the {\it ICO Prize in Optics} (2016), the inaugural {\it MDPI Materials Young Investigator Award} (2016), the {\it Kavli Foundation Early Career Lectureship in Materials Science} (2016), the inaugural {\it ACS Photonics Young Investigator Award Lectureship} (2016), the {\it Edith and Peter O'Donnell Award in Engineering} (2016), the {\it NSF Alan T. Waterman Award} (2015), the {\it IEEE MTT Outstanding Young Engineer Award} (2014), the {\it OSA Adolph Lomb Medal} (2013), the {\it IUPAP Young Scientist Prize in Optics} (2013), the {\it Franco Strazzabosco Award for Young Engineers} (2013), the {\it SPIE Early Career Investigator Award} (2012), the {\it URSI Issac Koga Gold Medal} (2011), an {\it NSF CAREER award} (2010), the {\it AFOSR} and the {\it DTRA Young Investigator Awards} (2010, 2011), {\it Young Scientist Awards} from {\it URSI General Assembly} (2005) and {\it URSI Commission B} (2010, 2007 and 2004). His students have also received several awards, including student paper awards at {\it IEEE Antennas and Propagation Symposia} (in 2011 to Y. Zhao, in 2012 to J. Soric). He is a {\it Simons Investigator in Physics} since 2016, has been selected twice as finalist of the {\it Blavatnik Award for Young Scientists} (2016, 2017), and is a {\it 2017 Highly Cited Researcher from Web of Science}. He has been serving as {\it OSA Traveling Lecturer} since 2010, {\it IEEE AP-S Distinguished Lecturer} since 2014, and as the {\it IEEE joint AP-S and MTT-S chapter for Central Texas}. Finally, he is a full member of {\it URSI}, a Fellow of {\it OSA}, {\it SPIE} and {\it APS}.
\end{IEEEbiography}

\begin{IEEEbiography}[{\includegraphics[width=1in,height=1.25in,clip,keepaspectratio]{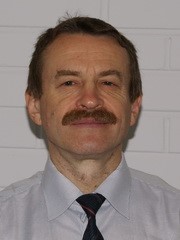}}]{Constantin (Konstantin) R. Simovski} worked in both industry and academic instituations in several countries, defended the PhD thesis in 1986 in the Leningrad Polytechincal Institute now St.-Petersburg Polytechnical University (Russia). In 2000 he defended in the same university the thesis of Doctor of Sciences in Physics and Mathematics (Habilitat, HDR). Full professor of ITMO University, St. Petersburg, Russia, in 2001-2008. Since 2008 he has been with Helsinki University of Technology, now – Aalto University. Full professor of Aalto University since 2012. Current research: metamaterials for optical sensing and energy harvesting, thermal radiation and radiative heat transfer on nanoscale, antennas for magnetic resonance imaging and wireless power transfer.
\end{IEEEbiography}

\begin{IEEEbiography}[{\includegraphics[width=1in,height=1.25in,clip,keepaspectratio]{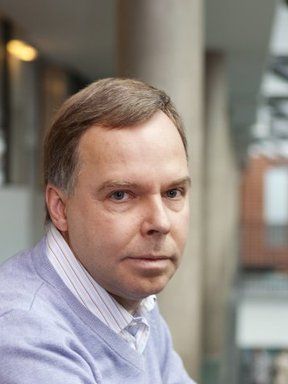}}]{Sergei A. Tretyakov} received the Dipl. Engineer-Physicist, the Candidate of Sciences (PhD), and the Doctor of Sciences degrees (all in radiophysics) from the St. Petersburg
State Technical University (Russia), in 1980, 1987, and 1995, respectively. From 1980 to
2000 he was with the Radiophysics Department of the St. Petersburg State Technical University. Presently, he is professor of radio science at the Department of Electronics and Nanoengineering, Aalto University, Finland. His main scientific interests are electromagnetic field theory, complex media electromagnetics, metamaterials, and
microwave engineering. He has authored or co-authored five research monographs and more than 280 journal papers. Prof. Tretyakov served as President of the Virtual Institute for Artificial Electromagnetic Materials and Metamaterials (``Metamorphose VI''), as General Chair, International Congress Series on Advanced Electromagnetic Materials in Microwaves and Optics (Metamaterials), from 2007 to 2013, and as Chairman of the St. Petersburg IEEE ED/MTT/AP Chapter from 1995 to 1998.
\end{IEEEbiography}

\end{document}